\documentclass[%
 reprint,
 superscriptaddress,
 amsmath,amssymb,
 aps,
 prx,
longbibliography 
]{revtex4-2}

\usepackage[pdftex]{graphicx} \graphicspath{{}}
\usepackage{braket}
\usepackage{bbm}
\usepackage{ulem}
\usepackage{verbatim}
\usepackage{bm}
\usepackage{mathtools}
\usepackage{xcolor}
\usepackage[pdftex,colorlinks=true]{hyperref}
\hypersetup{
  colorlinks   = true, 
  urlcolor     = blue, 
  linkcolor    = blue, 
  citecolor   = blue 
}

\usepackage{textcomp}

\usepackage[smallerops]{newtxmath}
\DeclareMathAlphabet{\mathcal}{OMS}{cmsy}{m}{n}

\newcommand{\pwc}[1]{{#1}}

\definecolor{C0}{HTML}{1F77B4}
\definecolor{C1}{HTML}{FF7F0E}
\definecolor{C2}{HTML}{2CA02C}
\definecolor{C3}{HTML}{D62728}
\definecolor{C4}{HTML}{9467BD}
\definecolor{C5}{HTML}{8C564B}
\definecolor{C6}{HTML}{E377C2}
\definecolor{C7}{HTML}{7F7F7f}
\definecolor{C8}{HTML}{BCBD22}
\definecolor{C9}{HTML}{17BECF}

\graphicspath{{./Figures/}}

\begin{document}

\title{Krylov complexity and Trotter transitions in unitary circuit dynamics}

\author{Philippe Suchsland}
\affiliation{Max Planck Institute for the Physics of Complex Systems, N\"{o}thnitzer Str. 38, 01187 Dresden, Germany}
\email{suchsland@pks.mpg.de}

\author{Roderich Moessner}
 \affiliation{Max Planck Institute for the Physics of Complex Systems, N\"{o}thnitzer Str. 38, 01187 Dresden, Germany}
 
\author{Pieter W. Claeys}
 \affiliation{Max Planck Institute for the Physics of Complex Systems, N\"{o}thnitzer Str. 38, 01187 Dresden, Germany}

\date{\today}

\begin{abstract}
\noindent
We investigate many-body dynamics where the evolution is governed by unitary circuits through the lens of `Krylov complexity', a recently proposed measure of complexity and quantum chaos. 
We extend the formalism of Krylov complexity to unitary circuit dynamics and focus on Floquet circuits arising as the Trotter decomposition of Hamiltonian dynamics. 
For short Trotter steps the results from Hamiltonian dynamics are recovered, whereas a large Trotter step results in different universal behavior characterized by the existence of local {\it maximally ergodic operators}: operators with vanishing autocorrelation functions, as exemplified in dual-unitary circuits.  
These operators exhibit maximal complexity growth, act as a memoryless bath for the dynamics, and can be directly probed in current quantum computing setups.
These two regimes are separated by a crossover in chaotic systems. Conversely, we find that free integrable systems exhibit a nonanalytic transition between these different regimes, where maximally ergodic operators appear at a critical Trotter step.
\end{abstract}

\maketitle

\section{Introduction}

One of the key goals of many-body physics is to understand and quantify the universal behavior in the dynamics of chaotic many-body systems. In the Heisenberg picture of quantum mechanics, all dynamics correspond to a unitary transformation of operators representing physical observables. If this dynamics is chaotic, initially ``simple'' operators are expected to become increasingly complex as time goes on. 
This complexity can, e.g., be observed through initially localized operators becoming increasingly delocalized, a process known as operator spreading, which is physically reflected in the growth of entanglement and operator scrambling. 
On the level of correlation functions, this increase of complexity results in all local information that is not encoded in conservation laws being lost, leading to correlation functions that eventually decay to universal (thermal) values. The intuition is that, by focusing on local observables, after sufficiently long times the remaining degrees of freedom are sufficiently complex that they can be treated as effectively random, allowing the rest of the system to be treated as a thermal bath \cite{dalessio_quantum_2016}.

Understanding and quantifying this increase of complexity is of both fundamental and practical importance. Fundamentally, it allows us to more precisely define notions of quantum chaos. On the practical level, it allows us to make predictions for the late-time dynamics of chaotic systems.

One possible way of quantifying the complexity of operator dynamics is through the so-called ``Krylov complexity''. Here, the dynamics of an initial operator is described in a Krylov subspace \cite{Viswanath1994}: the initial operator spreads out in a basis of operators obtained through a Gram-Schmidt orthonormalization procedure. 
The Krylov basis consists of orthonormal operators with increasing support and increasing complexity, acting as an effective bath for the initial operator (for a more precise definition, see the next section). 
This notion of complexity has recently been extensively studied for Hamiltonian dynamics, since it was conjectured that chaotic quantum dynamics results in a maximal growth of the Krylov complexity~\cite{parker_universal_2019}. The Krylov dynamics was also conjectured to satisfy a `universal operator growth hypothesis' and it was shown that Krylov complexity bounds a large class of physical complexity measures, including out-of-time-order correlators. 
It was subsequently realized that quantum chaos corresponds to delocalization in this Krylov subspace \cite{dymarsky_quantum_2020,rabinovici_krylov_2022,rabinovici_krylovlocalization_2022}. 

However, there has been increased interest in recent years in moving away from dynamics generated by a static local Hamiltonian, e.g., in driven systems \cite{bukov_universal_2015} or in unitary circuit models \cite{fisher_random_2023}. In such dynamics there is no longer a well-defined notion of a static Hamiltonian, conservation of energy does not hold, and the corresponding Krylov subspace can no longer be straightforwardly defined.

In this work we extend the framework of Krylov subspaces and the universal operator growth hypothesis from static Hamiltonians to unitary dynamics without an underlying local Hamiltonian, focusing on the dynamics described by unitary circuits. As one major motivation, this framework also allows us to study `Trotterized' circuits. These unitary circuits arise as the decomposition of Hamiltonian dynamics in discrete time steps \cite{trotter_product_1959,suzuki_relationship_1976}, and have both fundamental and practical applications: Trotterization is a key element in numerical algorithms for quantum many-body dynamics such as time-evolving block decimation (TEBD)~\cite{vidal_classical_2007,vidal_efficient_2004, White_2004,itensor}, and in the digital quantum simulation of quantum many-body dynamics \cite{lloyd_universal_1996}. This approach is more generally motivated by the enormous success that unitary circuits have had in the study of operator growth \cite{nahum_operator_2018,von_keyserlingk_operator_2018,khemani_operator_2018,chan_solution_2018,zhou_entanglement_2020,garratt_local_2021,fisher_random_2023,bertini_exact_2018,bertini_exact_2019,piroli_exact_2020,claeys_maximum_2020,claeys_ergodic_2021,aravinda_dual-unitary_2021,ho_exact_2022,borsi_construction_2022,claeys_emergent_2022,rather_construction_2022,rampp_dual_2023}. 
On the simplest level, unitary circuits provide minimally structured models mimicking local Hamiltonian dynamics, where the locality is directly encoded in the construction of the unitary evolution operator. 
Such circuits can either be purely random, where random unitary circuits have been highly successful in reproducing generic aspects of operator spreading \cite{nahum_operator_2018,von_keyserlingk_operator_2018,khemani_operator_2018,chan_solution_2018,zhou_entanglement_2020,garratt_local_2021,fisher_random_2023}, or structured, where e.g., dual-unitary circuits have allowed for exact results on many-body dynamics and spectral measures of quantum chaos \cite{bertini_exact_2018,bertini_exact_2019,piroli_exact_2020,claeys_maximum_2020,claeys_ergodic_2021,aravinda_dual-unitary_2021,ho_exact_2022,borsi_construction_2022,claeys_emergent_2022,rather_construction_2022,rampp_dual_2023}.
Remarkably, various Trotterized circuits arising from integrable Hamiltonians remain integrable at any Trotter step, and both noninteracting and interacting integrable circuit models exist \cite{vanicat_integrable_2018,ljubotina_ballistic_2019,
friedman_spectral_2019,gombor_integrable_2021,sa_integrable_2021,sopena_algebraic_2022,giudice_temporal_2022,
claeys_correlations_2022,miao_floquet_2022,vernier_integrable_2022,miao_integrable_2023}. 

We first argue that, for generic unitary dynamics, the dynamics in the Krylov subspace approaches a universal form again characterized by a maximal growth of the Krylov complexity. This maximal growth is accompanied by the convergence of the Krylov basis operators to `maximally ergodic operators': operators for which the autocorrelation function vanishes instantaneously. In this way these Krylov basis operators act as a memoryless bath. 
For these operators the usual description of many-body dynamics in terms of the eigenstate thermalization hypothesis (ETH) reduces to a purely random matrix theory (RMT) prediction, highlighting the effect of the Trotterization and the deviation from Hamiltonian dynamics. 
These operators act as attractive points in the Krylov subspace and are reminiscent of classical Bernoulli systems \cite{ott_chaos_2002}, which are similarly maximally ergodic. These maximally ergodic operators present a new intuition for the relaxation to equilibrium in unitary many-body dynamics, and their dynamics can be directly probed in current quantum computing setups \pwc{using recent proposals for the experimental measurement of (infinite-temperature) autocorrelation functions~\cite{richter_simulating_2021,Mi_2021}}. %{\color{blue}(Appendix~\ref{sec:app_correlation_function_measurement})}. 

For chaotic systems, we argue that with increasing Trotter step there is a crossover in the Krylov dynamics from the Hamiltonian regime, where the dynamics satisfy the universal operator growth hypothesis, to the maximally ergodic ergime, where the Krylov operators converge to maximally ergodic operators. In the latter regime the dynamics can no longer be accurately described using a local Hamiltonian. The fundamental implication is that, for any nonzero Trotter step, it is possible to construct large but local operators for which the dynamics reduces to the purely random matrix prediction and the autocorrelation function trivializes, no matter how close the circuit appears to be to Hamiltonian dynamics.

In noninteracting integrable circuits we observe a nonanalytic Trotter transition rather than a crossover between different regimes as the Trotter step is varied. For sufficiently small Trotter step no maximally ergodic operators exist and no maximal Krylov complexity growth is possible, indicating that the possible dynamics remains highly constrained by the conservation laws, as in the Hamiltonian case, but at a critical Trotter step maximally ergodic operators appear as attractors in the Krylov subspace and the Krylov dynamics approaches this universal regime. Remarkably, in this regime the autocorrelation function can appear maximally ergodic despite the highly nonergodic nature of the underlying circuit.

This behavior can be understood through the spectral function of the Krylov operators, i.e., the Fourier transform of their autocorrelation function, which determines their response to external perturbations. The Gram-Schmidt procedure used to construct the Krylov basis results in basis operators whose spectral function is increasingly flattened and the correlation functions increasingly featureless, indicating that the dynamics converges to operators with a flat spectral function and a corresponding instantaneous decay of correlation functions. 
Crucially, the spectral functions of the Krylov operators are proportional to the spectral function of the initial operator, explaining the nonanalytical transition in free systems: for sufficiently small Trotter steps the spectral function has a finite support, scaling linearly with the Trotter step, and a resulting gap in the excitation spectrum (around the $\pi$-mode of the Floquet Brillouin zone), but at the critical Trotter step this gap closes. A closed gap is necessary in order to have a flat spectrum for the Krylov operators, whereas in the presence of a gap no maximally ergodic operators can appear, since this gap is inherited by the spectral functions of all Krylov operators. 

While this effect might seem surprising, there is a class of circuits for which it already has been shown that maximally ergodic operators exist: dual-unitary circuits, characterized by an underlying space-time duality \cite{gopalakrishnan_unitary_2019,bertini_exact_2019,claeys_ergodic_2021}. This space-time duality effectively enforces the Trotter step to be large, and dual-unitary circuits can be either integrable or chaotic while still supporting maximally ergodic dynamics. For the Trotterized circuits considered here, dual-unitary gates appear deep in the maximally ergodic regime.

This paper is structured as follows. The full formalism of Krylov dynamics is presented in Sec.~\ref{sec:formalism} and is subsequently applied to dual-unitary circuits in Sec.~\ref{sec:dualunitary}, both as an illustrative example for which analytical calculations are possible, and to illustrate the existence of maximally ergodic operators as attractors in the Krylov subspace. In Sec.~\ref{sec:trotterized} the formalism is applied to Trotterized circuits and numerical results are presented for chaotic, interacting integrable, and noninteracting integrable circuits. The nonanalytic Trotter transition in the latter case is subsequently discussed in more detail in Sec.~\ref{sec:noninteracting}, where we present analytic results in a representative example. \pwc{The growth of Krylov complexity in these different classes of circuits is illustrated in Sec.~\ref{sec:complexity}.}

We note that the idea of using Krylov subspaces to study unitary dynamics was already used in Refs.~\cite{yates_strong_2021,yates_edge_2022} to study weak and strong zero modes in Floquet spin chains. Our approach differs in that we make extensive use of the unitarity to restrict our theoretical framework, such that we will not need to consider the full unitary superoperator, and our focus on unitary circuits, which allows us to make statements about the locality of Krylov operators and make the connection with Hamiltonian dynamics \cite{parker_universal_2019}.

\section{Formalism}
\label{sec:formalism}
In this work we consider unitary evolution described by so-called `brickwork' circuits, with the added restriction that we require the dynamics to be periodic in time (leading to Floquet dynamics \cite{bukov_universal_2015}). Such brickwork circuits describe the dynamics of an infinite one-dimensional lattice in a discrete time, and are constructed from a local two-site unitary gate. Taking $U_{j,j+1}$ to be a two-site unitary matrix acting on sites $j$ and $j+1$, we consider a unitary evolution operator for a single discrete time step as 
\begin{equation}\label{eq:U_tot}
U = U_{\textrm{odd}}U_{\textrm{even}},
\end{equation}
with
\begin{align}
U_{\textrm{odd}} = \prod_{j \in \mathbb{Z}}U_{2j-1,2j}, \quad U_{\textrm{even}} = \prod_{j \in \mathbb{Z}} U_{2j,2j+1}\,.
\end{align} 
The resulting unitary dynamics can be graphically represented in tensor network language as
\begin{align}\label{eq:brickwork}
U^t =\,\, \vcenter{\hbox{\includegraphics[width=0.6\linewidth]{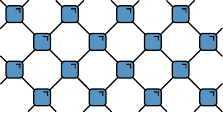}}}
\end{align}
here illustrated for two time steps, $t=2$, and eight lattice sites, and where every blue square represents a unitary gate acting on two sites. For convenience we take all unitary gates to be identical, although this is not a necessary assumption for our framework. Note that a single time step corresponds to two rows of unitary gates, such that the dynamics is periodic with period one. 

The main focus of this work is on describing the operator dynamics generated by such a unitary circuit in a Krylov subspace. 
While the developed formalism will be applicable to more general unitary dynamics, we focus on unitary circuits because they allow for a direct connection with previous results on Krylov dynamics. 
As the space of all operators is a Hilbert space itself, we use the Dirac notation for the operator space and write $|O)\equiv O$ to represent the operator $O$. Here we use round brackets to refer to the operator Hilbert space, in contrast to the commonly used angle brackets denoting the Hilbert space of states. The scalar product is defined as $(O_1|O_2) = \mathrm{Tr}[O_1^\dagger O_2]/\mathcal{D}$, where $\mathcal{D}$ is the dimension of the (state) Hilbert space. In the remainder of this work we will typically work in the thermodynamic limit of an infinite system size.

Any operator $O$ evolves in discrete time under the repeated application of the unitary transformation $U$, corresponding to, e.g., a Floquet unitary or a single time step in a unitary brickwork circuit. We define the unitary superoperator $\mathcal{U}$ to act as $\mathcal{U}|O) = |U^{\dagger}OU)$ and consider an initial operator $O_0$ which we take to be Hermitian, traceless, and normalized as $(O_0|O_0)=1$. The full dynamics is encoded in the sequence of operators
\begin{equation}
\left|O_t\right) = \mathcal{U}^t \left|O_0\right), \quad t \in \mathbb{N}_0\,,
\end{equation}
giving the time-evolved operator at every discrete time $t$. 
The autocorrelation function for $O_0$ follows directly as
\begin{equation}
(O_0|O_t) = (O_0|\,\mathcal{U}^t|O_0) = \mathrm{Tr}\left[O_0^\dagger U^{\dagger t} O_0 U^t\right]/\mathcal{D}\,. \label{eq:autocorrf}
\end{equation}
The Krylov subspace of an initial operator $O_0$ is spanned by basis elements obtained from an appropriate orthonormalization of the sequence $\{|O_0),|O_1) ,|O_2)  \dots\}$.

If the initial operator $O_0$ is initially localized, e.g., on a single site, the support of the operator $|O_t)$ grows linearly with $t$ due to the unitarity and the brickwork geometry \cite{chan_solution_2018,claeys_maximum_2020}. Using the unitarity of the individual gates, a single-site operator can be directly shown to evolve as
\begin{align}
\vcenter{\hbox{\includegraphics[width=0.7\linewidth]{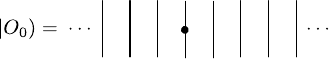}}}\,
\end{align}
\begin{align}
\vcenter{\hbox{\includegraphics[width=0.7\linewidth]{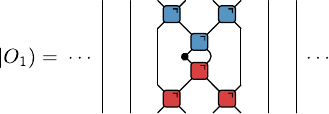}}}\,
\end{align}
\begin{align}\label{eq:brickwork_support}
\vcenter{\hbox{\includegraphics[width=0.7\linewidth]{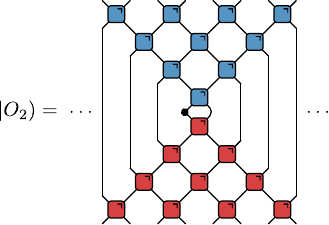}}}\,
\end{align}
where the black circle denotes the initial operator and the red squares represent the Hermitian conjugate of the unitary gates. This linear growth then bounds the support of the Krylov operators. 

Written in this way, the operator dynamics has been extensively analyzed in, e.g., random unitary circuits \cite{nahum_operator_2018,von_keyserlingk_operator_2018,khemani_operator_2018,chan_solution_2018,zhou_entanglement_2020,garratt_local_2021,fisher_random_2023}. Here, we take a different approach and consider the operator dynamics in an orthonormal basis, obtained through a Gram-Schmidt orthonormalization of the set of operators $|O_t)$. This approach is also known as the Arnoldi iteration \cite{Arnoldi1951}.

\subsection{Hermitian superoperator}
The Krylov approach has been instructive when considering Hamiltonian dynamics, where the operator dynamics is governed by a Liouvillian $\mathcal{L}(O) = [H,O]$ for a given Hamiltonian $H$. In this case, the orthonormalization is done on a set of operators $|O_n) = \mathcal{L}^n |O_0)$. We briefly review the main formalism from Ref.~\cite{parker_universal_2019}.

Identifying $|\mathcal{O}_0) = |O_0)$ and $|\mathcal{O}_1) = \mathcal{L}|\mathcal{O}_0) /\tilde{b}_1$ with $\tilde{b}_1 = (\mathcal{L}\mathcal{O}_0|\mathcal{L}\mathcal{O}_0)^{1/2}$, it is possible to inductively define
\begin{align}
&|A_n )  = \mathcal{L}|\mathcal{O}_{n-1}) - \tilde{b}_{n-1}|\mathcal{O}_{n-2}),\\
&\tilde{b}_n = (A_n|A_n)^{1/2},\quad |\mathcal{O}_n) = |A_n)/\tilde{b}_n\,.
\end{align}
These operators form an orthonormal basis for the operator dynamics, $(\mathcal{O}_m|\mathcal{O}_n) = \delta_{mn}$, and arise as a Gram-Schmidt orthonormalization of the sequence $\{\mathcal{L}^n |O_0)\}$. The Liouvillian is tridiagonal in this basis, 
\begin{align}\label{eq:Lmn}
\mathcal{L}_{mn} = (\mathcal{O}_m|\mathcal{L}|\mathcal{O}_n) = 
\begin{pmatrix}
0 & \tilde{b}_1 & 0  & 0 & \dots \\
\tilde{b}_1 & 0 & \tilde{b}_2 & 0 & \dots \\
0 & \tilde{b}_2 & 0 & \tilde{b}_3 & \dots \\
0 & 0 & \tilde{b}_3 & 0 & \dots \\
\vdots & \vdots & \vdots & \vdots & \ddots
\end{pmatrix}\,.
\end{align}
The tridiagonal form is direct consequence of the Hermiticity of the superoperator $i\mathcal{L}$, and the diagonal elements vanish because of the commutator structure of $\mathcal{L}$. The Lanczos coefficients $\{\tilde{b}_n\}$ fully encode the Liouvillian in this basis. A review of this recursion approach can be found in Ref.~\cite{Viswanath1994}, and the Lanczos coefficients were argued to exhibit universal behavior in Ref.~\cite{parker_universal_2019}. Specifically, for a nonintegrable lattice Hamiltonian the Lanczos coefficients were argued to asymptotically grow linearly with $n$, $\tilde{b}_n \propto n$, up to a logarithmic correction in one-dimensional systems. This constitutes the ``universal operator growth hypothesis'' of Ref.~\cite{parker_universal_2019}. This linear growth was contrasted with the boundedness of the Lanczos coefficients in noninteracting integrable systems,  $\tilde{b}_n \propto \textrm{Cst.}$, and the sublinear growth observed in interacting integrable systems, where for the integrable XXZ Hamiltonian a growth $\tilde{b}_n \propto \sqrt{n}$ was observed.

There have subsequently been a large number of recent studies on Krylov dynamics ranging from spin models \cite{noh_operator_2021,ballar_trigueros_krylov_2022,heveling_numerically_2022,Bhattacharjee_2022,qi_surprises_2023,yeh_slowly_2023}, models with restricted symmetry group \cite{rabinovici_operator_2021,baek_krylov_2022,patramanis_probing_2022,chapman_unified_2023,iizuka_krylov_2023}, quantum scars \cite{moudgalya_hilbert_2022,moudgalya_symmetries_2022,Bhattacharjee2022}, billiards \cite{hashimoto_krylov_2023} and quantum networks \cite{kim_operator_2022}, to (holographic) field theories \cite{jian_complexity_2021,caputa_operator_2021,dymarsky_krylov_2021,adhikari_cosmological_2022,kar_random_2022,avdoshkin_krylov_2022,banerjee_cfts_2022,camargo_krylov_2022}.
The associated notion of Krylov complexity has also been shown to be intimately related to various notions of entanglement and complexity \cite{caputa_geometry_2022,balasubramanian_quantum_2022,fan_growth_2022,fan_universal_2022,lv_building_2023} as well as physical observables \cite{Pal2023,zhang_universal_2023} and satisfies a quantum speed limit \cite{hornedal_geometric_2023}. Krylov methods have also found natural applications in counterdiabatic driving \cite{claeys_floquet-engineering_2019,takahashi_shortcuts_2023,bhattacharjee_lanczos_2023}.

\subsection{Unitary superoperator}
For the unitary superoperator $\mathcal{U}$ relevant for this work both the orthonormalization procedure and the resulting matrix representation of the superoperator are slightly more involved. A detailed derivation can be found in Appendix \ref{sec:app_superoperator_structure}, and we here only quote the final result. The Krylov operators $\{|\mathcal{O}_0),|\mathcal{O}_1), |\mathcal{O}_2),\dots\}$ are obtained through a Gram-Schmidt orthonormalization of the operators $\{|O_0),|O_1), |O_2),\dots\}$, satisfying $(\mathcal{O}_m|\mathcal{O}_n)=\delta_{mn}$. The Krylov basis can be expanded as
\begin{align}\label{eq:decompose_ev_op_krylov}
|\mathcal{O}_n) = \sum_{t=0}^n \alpha_{n,t}|O_t), 
\end{align}
which can be inverted as
\begin{align}\label{eq:invert_transfo_krylov}
|O_t) = \sum_{n=0}^t \beta_{t,n}|\mathcal{O}_n).
\end{align}
If the initial operator $|O_0)$ is local, the operators $|\mathcal{O}_n)$ are similarly local, with a support that grows linearly in $n$ \cite{chan_solution_2018,claeys_maximum_2020}. In this basis $\mathcal{U}$ has the structure of an upper Hessenberg form, whose matrix elements $\mathcal{U}_{mn}$ vanish for $m>n+1$, as opposed to the tridiagonal matrix observed in the Hermitian case (as also observed in Ref.~\cite{yates_strong_2021} and for open systems in Ref.~\cite{Bhattacharya_2022}). Furthermore, the unitary superoperator can be fully parametrized by three functions:
\begin{align}\label{eq:U_nm}
&\mathcal{U}_{mn} = (\mathcal{O}_m|\,\mathcal{U}|\mathcal{O}_n) =\begin{cases}
0 \quad &\textrm{if} \quad m>n+1\,,\\
b_m \quad &\textrm{if} \quad m = n+1\,,\\
a_m\, c_n/c_m \quad &\textrm{if} \quad m<n+1\,,
\end{cases}
\end{align}
leading to an upper Hessenberg matrix of the form 
\begin{align}\label{eq:U_Hess}
\mathcal{U}_{mn} =
\begin{pmatrix}
a_0 && a_0 c_1/c_0 && a_0 c_2/c_0 && a_0  c_3/c_0 & \dots \\
b_1 && a_1  &&  a_1 c_2/c_1 && a_1 c_3/c_1 & \dots \\
0 &&  b_2 && a_2  &&  a_2 c_3/c_2 & \dots \\
0 && 0 &&  b_3 && a_3  &\dots \\
0 && 0 &&  0 && b_4 &\dots \\
\vdots && \vdots && \vdots && \vdots &  \ddots 
\end{pmatrix}\,.
\end{align}
The three sequences correspond to specific matrix elements 
\begin{align}
&a_n = (\mathcal{O}_n|\,\mathcal{U}|\mathcal{O}_{n}), \\
&b_n = (\mathcal{O}_{n}|\,\mathcal{U}|\mathcal{O}_{n-1}), \\
&c_n =  (\mathcal{O}_0|\,\mathcal{U}|\mathcal{O}_{n}).
\end{align}
Note that $a_0=c_0$. 
The unitarity further constrains these matrix elements in various ways, e.g., $|\,\mathcal{U}_{mn}| \leq 1, \forall m,n$, which already indicates that the unbounded growth of the Lanczos coefficients observed in Hamiltonian dynamics should not be possible in unitary circuit dynamics. In general we observe $c_n$ to be a decaying sequence, such that the matrix is dominated by the diagonal elements. The structure of $b_n$ and $a_n$ appears to be universal for different classes of systems, in a way that will be made explicit later, with $b_n \to 1$ and $a_n \to 0$ for sufficiently large $n$.

\begin{figure}[htb]
    \centering
    \includegraphics[width=0.7\columnwidth]{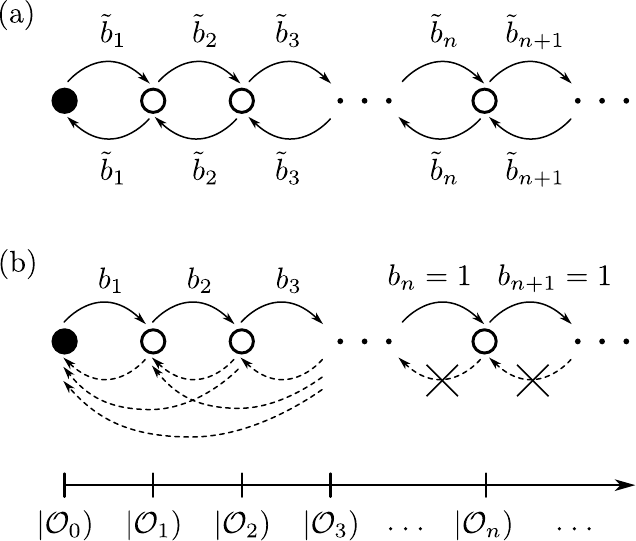}
    \caption{Representation of operator growth as dynamics on a semi-infinite one-dimensional chain in the Krylov subspace. (a) For Hamiltonian dynamics the Krylov dynamics reduces to a nearest-neighbor hopping model with hopping coefficients $\tilde{b}_n$. (b) For unitary circuit dynamics the hopping is asymmetric, with nearest-neighbor hopping to the right with coefficients $b_n$ and long-range hopping terms to the left set by the coefficients $a_n$ and $c_n$. However, for sufficiently large $n$ the hopping to the right generally dominates and the hopping strength approaches $b_n=1$, whereas the hopping terms to the left vanish. 
    \label{fig:hopping}}
\end{figure}

The matrix representations \eqref{eq:Lmn} and \eqref{eq:U_Hess} allow for an interpretation of the operator dynamics as dynamics in a semi-infinite one-dimensional chain where each lattice site $n$ represents a Krylov operator $\mathcal{O}_n$ and the initial operator is initially localized at site $n=0$. The different hopping models for Hamiltonian and unitary dynamics are illustrated in Fig.~\ref{fig:hopping}. 

Recasting the dynamics in a Krylov subspace allows for the evaluation of the so-called Krylov complexity, also known as K-complexity, which measures the spreading of an operator in Krylov space. Following Eq.~\eqref{eq:invert_transfo_krylov}, we define the Krylov complexity for unitary circuit dynamics as
\begin{align} \label{eq:our_def_krylov_compl}
K(t) = \sum_{n=0}^t n |\beta_{t,n}|^2 \,.
\end{align}
The Krylov complexity was originally proposed as a measure of chaos for continuous time evolution \cite{parker_universal_2019}. In this work we extend this approach to Floquet evolution and study its universal properties.

\subsection{Spectral functions}
\label{subsec:spectral}
\pwc{In this subsection we present an alternative interpretation of the orthonormalization procedure, which will be important in understanding the autocorrelation functions of Krylov operators and Trotter transitions. Readers mainly interested in the application of the developed framework to specific unitary circuits can directly skip to the next section.}
The action of the orthonormalization procedure can be reinterpreted on the level of the spectral function, also known as the structure function,
\begin{equation}\label{eq:spectral_function_for_operator}
|f_O(\omega)|^2 = \frac{1}{\mathcal{D}}\sum_{p,q=1}^{\mathcal{D}}|\langle p|O|q \rangle|^2\, \delta(\theta_q-\theta_p-\omega)\,,
\end{equation}
where $U \ket{p} = e^{i\theta_p}\ket{p}$. The spectral function is the Fourier transform of the autocorrelation function
\begin{equation}
\frac{1}{\mathcal{D}}\mathrm{Tr}\left[O^\dagger U^{\dagger t} O U^t \right] = \frac{1}{\mathcal{D}}\sum_{p,q=1}^{\mathcal{D}}|\langle p|O|q \rangle|^2 \, e^{i(\theta_q-\theta_p)t}\,.
\end{equation}
As such, it encodes information about both the short- and long-time dynamics of the initial operator and hence the level of ergodicity. Within ETH, the spectral function naturally appears as the (averaged over eigenstates) envelope for the off-diagonal matrix elements \cite{dalessio_quantum_2016}. For unitary dynamics where energy is no longer conserved, ETH predicts a universal form for the off-diagonal elements of an observable $O$ as
\begin{equation}
\braket{p|O|q} =  \mathcal{D}^{-1/2} f_O(\theta_q-\theta_p)R_{pq},
\end{equation}
where $R_{pq}$ is a random variable with zero mean and unit variance~\cite{dalessio_long-time_2014,lazarides_equilibrium_2014}. The operator dynamics is encoded here in the spectral function appearing as an envelope for off-diagonal matrix elements, distinguishing structured dynamics from those generated by a purely random matrix.  
However, the spectral function  remains well-defined through the autocorrelation function even when ETH is not expected to hold and can be used as a sensitive probe of integrability and chaos \cite{pandey_adiabatic_2020,brenes_eigenstate_2020,leblond_universality_2021}. 

The orthonormality of the Krylov operators can be directly translated to statements about the orthonormality of polynomials on the unit circle, which can in turn be used to make exact statements about the action of the repeated orthonormalization. Such a connection was already made for Hamiltonian dynamics in Ref.~\cite{muck_krylov_2022}, but is simpler for unitary dynamics and allows us to make use of exact statements for orthonormal polynomials on the unit circle that do not apply for orthonormal polynomials on the real line (as in the  Hamiltonian case). Expanding the Krylov operators in the original basis \pwc{ as in Eq.~\eqref{eq:decompose_ev_op_krylov},}
\begin{equation} \label{eq:repeat_decomp}
\left|\mathcal{O}_n\right) = \sum_{t=0}^n \alpha_{n,t} \left|O_t\right)\,,
\end{equation}
we define corresponding functions on the unit circle as
\begin{equation}\label{eq:def_orthpoly}
p_n(\omega) = \sum_{t=0}^n \alpha_{n,t}\, e^{it\omega}\,.
\end{equation}
These functions now behave as orthonormal functions provided we use the spectral function as weight function,
\begin{equation}
\int_0^{2\pi}\!\mathrm{d}\omega\, |f_O(\omega)|^2 p_n(\omega)p_m^*(\omega) = \delta_{mn}\,.
\end{equation}
This orthonormality directly follows from a spectral expansion of the orthonormality of the Krylov operators, since
\begin{align}
&\mathrm{Tr}\left[\mathcal{O}_m^{\dagger}\mathcal{O}_n \right] / \mathcal{D} = \sum_{t=0}^m \sum_{s=0}^n \alpha^*_{m,
t} \alpha_{n,s}\mathrm{Tr}\left[O_{t}^{\dagger} O_s \right]/\mathcal{D} \nonumber\\
&\qquad=  \sum_{t=0}^m \sum_{s=0}^n \alpha^*_{m,
t}\alpha_{n,s}\int_0^{2\pi}\!\mathrm{d}\omega\, |f_O(\omega)|^2 e^{i \omega(s-t)} \nonumber \\
&\qquad = \int_0^{2\pi}\!\mathrm{d}\omega\, |f_O(\omega)|^2 p_n(\omega)p_m^*(\omega)\,.
\end{align}

These orthonormal functions are not just a mathematical curiosity. Rather, they fully determine the autocorrelation functions and associated spectral functions of the Krylov operators themselves. The corresponding spectral function can be directly calculated as
\begin{align}\label{eq:spectral_On}
&|f_{\mathcal{O}_n}(\omega)|^2  = \frac{1}{\mathcal{D}}\sum_{p,q=1}^{\mathcal{D}}|\langle p|\mathcal{O}_n|q \rangle|^2\, \delta(\theta_q-\theta_p-\omega) \nonumber \\
&\qquad=\frac{1}{\mathcal{D}}\sum_{p,q=1}^{\mathcal{D}}|\langle p|\mathcal{O}_0|q \rangle|^2  |p_n(\theta_q-\theta_p)|^2 \, \delta(\theta_q-\theta_p-\omega)  \nonumber\\
&\qquad=|f_O(\omega)|^2 |p_n(\omega)|^2\,.
\end{align}
%{\color{blue} which follows from the definitions of $O_t,\, |p\rangle,\, p_n,\, f_{\mathcal{O}_n}(\omega)$ and Eq.~\eqref{eq:repeat_decomp}}.
%%
\pwc{Here we have used that
\begin{align}
    \langle p|\mathcal{O}_n|q\rangle = \langle p|\mathcal{O}_0|q\rangle p_n(\theta_q - \theta_p),
\end{align}
as follows directly from the definition of $p_n(\omega)$ and Eq.~\eqref{eq:repeat_decomp}:
\begin{align}
    &\langle p|\mathcal{O}_n|q\rangle = \sum_{t=0}^{n} \alpha_{n,t} \langle p| O_t|q\rangle= \sum_{t=0}^{n} \alpha_{n,t} \langle p| U^{\dagger t} O_0 U^t|q\rangle
 \nonumber\\
 &\quad= \sum_{t=0}^{n} \alpha_{n,t}e^{it(\theta_q - \theta_p)} \langle p| O_0|q\rangle =   p_n(\theta_q - \theta_p) \langle p| O_0|q\rangle\,.
\end{align}
}
\pwc{We can now use these orthonormal polynomials to derive a result on the spectral function of Krylov operators. As shown in Refs.~\cite{szego_orthogonal_1975,nevai_geza_1986}}, such orthonormal polynomials have the property that the first $n$ Fourier modes of $|p_n(\omega)|^{-2}$ agree with the first $n$ Fourier modes of $|f_O(\omega)|^{2}$, i.e.,
\begin{align}
    \int_0^{2\pi} \!\mathrm{d}\omega\, e^{im \omega} |p_n(\omega)|^{-2} &= \int_0^{2\pi} \!\mathrm{d}\omega\, e^{im \omega} |f_O(\omega)|^2, \nonumber \\
    &\qquad \textrm{if} \quad |m| \leq n\,.
\end{align}
These polynomials hence satisfy
\begin{align}\label{eq:convergence_pn_0}
\lim_{n \to \infty}|p_n(\omega)|^{-2} = \frac{1}{2\pi}|f_O(\omega)|^{2}\,,
\end{align}
provided $\log(|f_O(\omega)|^2)$ is Lebesgue integrable \cite{szego_orthogonal_1975,nevai_geza_1986}. This convergence will be important when considering the large-$n$ limit of Krylov operators, since it implies that their spectral function generally converges to a constant. Additional properties of such orthonormal polynomials on the unit circle and their relation to the Fourier modes of the spectral function are reviewed in Appendix~\ref{app:orthonormalpoly}.

\section{Dual-unitary circuits}
\label{sec:dualunitary}
Let us illustrate the theoretical framework from the previous section for the specific case of dual-unitary circuits. These circuits are characterized by an underlying space-time duality, allowing their (auto-)correlation functions to be calculated exactly, and support both chaotic and integrable dynamics \cite{gopalakrishnan_unitary_2019,bertini_exact_2019,claeys_ergodic_2021}. As such, they also provide a tractable case for studying Krylov complexity in Floquet circuits, that will be instructive when considering more complicated circuit dynamics. \pwc{In the following we will consider a random dual-unitary circuit, taking all gates $U_{2j-1,2j}$ and $U_{2j,2j+1}$ in Eq.~\eqref{eq:U_tot} to be identical dual-unitary gates constructed using the parametrization from Refs.~\cite{bertini_exact_2019,claeys_correlations_2022} with a random choice of parameters.} 

\subsection{Local operators}
In dual-unitary circuits all correlation functions for single-site operators vanish everywhere except on the edge of the causal light cone ~\cite{bertini_exact_2019,claeys_ergodic_2021}. As one immediate consequence, the autocorrelation vanishes at all times except for $t=0$ in dual-unitary circuit dynamics, i.e.,
\begin{align}
(O_0|O_t) = \delta_{t,0},
\end{align}
provided we choose $O_0$ as an single-site operator, e.g., $O_0 = \sigma_{j}$ only acting nontrivially on a fixed site $j$. Combining this result with $(O_{s}|O_t)=(O_0|O_{t-s})$, we find that the Krylov operators can be identified with the (normalized) time-evolved operators $|\mathcal{O}_n) = |O_n)$. The unitary superoperator takes a particularly simple form since $\mathcal{U}|\mathcal{O}_n) = |\mathcal{O}_{n+1})$ and we can write
\begin{equation}\label{eq:DU_Krylov}
\mathcal{U} = \begin{pmatrix}
0 & 0 & 0  & 0 & \dots \\
1 & 0 & 0 & 0 & \dots \\
0 & 1 & 0 & 0 & \dots \\
0 & 0 & 1 & 0 & \dots \\
\vdots & \vdots & \vdots & \vdots & \ddots
\end{pmatrix}.\,
\end{equation}
In the language of Eq.~\eqref{eq:U_nm}, we find that $b_n=1, \forall n$ and $a_n = 0, \forall n$. Every operator generated by the unitary superoperator is orthogonal to all previous ones. The Krylov complexity grows linearly in time, and we find that $K(t) = t$ since $\alpha_{n,t} = \beta_{t,n} = \delta_{t,n}$ for $|\mathcal{O}_n) = |O_n)$. Such a growth is in fact the maximal possible growth due to the unitarity of the superoperator. \pwc{ The unitarity restricts $|b_n| \leq 1$, and these coefficients determine the rate of hopping to the right, such that the growth of $K(t)$ is maximal for $|b_n|=1$ taking its maximal value and correspondingly $|a_n|=0$ its minimal value, see also Appendix \ref{sec:app_superoperator_structure}.}

Furthermore, since the autocorrelation function is a delta function, the spectral function is a constant,
\begin{align}
|f_{O_t}(\omega)|^2 = \frac{1}{2\pi}\,.
\end{align}
\begin{figure}
    \centering
    \includegraphics[width=\columnwidth]{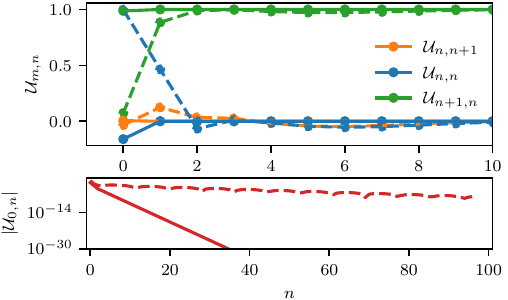}
    \caption{
    The unitary superoperator $\mathcal{U}$ [Eq.~\eqref{eq:U_nm}] describing the evolution of the initial operator $|O_0) \propto \sum_{j \in \mathbb{Z}} \sigma^z_{2j}$ under dual-unitary dynamics. Here $\mathcal{U}$ is given in the Krylov basis, which is generated by orthonormalization of $\{\mathcal{U}^t|O_0)\}$ with $t\in \mathbb{N}_0$. The unitary superoperator $\mathcal{U}$ is fully described by its tridiagonal matrix elements (top) and its decay away from the diagonal $\mathcal{U}_{n,n+l}=\mathcal{U}_{n,n+1}\mathcal{U}_{0,n+l}/\mathcal{U}_{0,n+1}$ (bottom). We consider two exemplary realisations of the dual-unitary gate: a typical one (solid) and one with slow convergence (dashed). After an initial transient behavior the coefficients $a_n = \mathcal{U}_{n,n}$ and $b_n = \mathcal{U}_{n+1,n}$ approach $0$ and $1$ respectively (top) and the coefficients $c_n = \mathcal{U}_{0,n}$ decay exponentially (bottom), indicating that the unitary superoperator becomes dominated by the tridiagonal elements. 
    \label{fig:tridiagonal_du}}
\end{figure}
\subsection{Sums of local operators}
A less trivial example can be found by considering an initial operator that is a linear combination of local operators, \pwc{ $O_0 \propto \sum_{j \in \mathbb{Z}} \sigma_{2j}$ with $\sigma_{2j}$ a single site operator}. This setup is similar to the one considered in Ref.~\cite{fritzsch_ETH_2021}, probing the spectral function in dual-unitary circuits. As detailed in Refs.~\cite{bertini_exact_2019,fritzsch_ETH_2021}, the autocorrelation functions now take the form
\begin{equation}
(O_0|O_t) = \mathrm{Tr}\left[\sigma \mathcal{M}^{2t}(\sigma)\right],
\end{equation}
where $\mathcal{M}$ is a quantum channel acting as
\begin{equation}\label{eq:channel}
\mathcal{M}(\sigma) = \frac{1}{q}\mathrm{Tr}_1\left[\tilde{U}^{\dagger}(\sigma \otimes \mathbbm{1})\tilde{U}\right]\,.
\end{equation}
\pwc{Here $\tilde{U} \equiv U_{j,j+1}, \forall j$ corresponds to the choice of dual-unitary gate and $q$ is the dimension of the local Hilbert space, which we take to be $q=2$ in this work.} Both $\tilde{U}$ and $\sigma \otimes \mathbbm{1}$ act on two copies of the local Hilbert space and $\mathrm{Tr}_1$ indicates that we are tracing out the first copy.
The autocorrelation function can be directly found through an exact diagonalization of the quantum channel $\mathcal{M}$. Introducing an eigenvalue decomposition of $\mathcal{M}$ as
\begin{align}
\mathcal{M} = \sum_{a=1}^{q^2} \lambda_a\, |r_a)(l_a|,
\end{align}
the autocorrelation function follows as
\begin{align} \label{eq:autocorr_du}
(O_0|O_t) = \sum_{a=1}^{q^2} \lambda_a^{2t}\, (\sigma|r_a)(l_a|\sigma),
\end{align}
from which the Krylov operators can be directly calculated \pwc{since all necessary overlaps in the orthonormalization procedure are of the form $(O_{s}|O_t) = (O_{t}|O_s)^* = (O_{0}|O_{t-s})$ for $t\geq s$}. The resulting matrix elements are illustrated in Fig.~\ref{fig:tridiagonal_du} for two generic choices of dual-unitary circuit and with $\sigma=\sigma_z$ a Pauli matrix for $q=2$. After an initial transient behavior for small $n$, the matrix elements approach the same form for large $n$ as previously observed, where only $b_n=1$ is nonzero. We now find that the Krylov complexity grows linearly after some time, with $K(t) = t + \textrm{Cst}$. 

This behavior can be understood by considering the case where our initial operator is an eigenoperator of the quantum channel \eqref{eq:channel}, yielding $(O_{n+k}|O_n) = \lambda^{2k}$~\footnote{We here assume that $\lambda$ is real, which is necessary if the corresponding eigenoperator is Hermitian.}. Alternatively, this expression for the autocorrelation function can be understood as an approximation for the late-time dynamics if these are dominated by a single eigenmode of the quantum channel. The Krylov operators can then be analytically calculated as
\begin{align}\label{eq:DU_Krylovop}
|\mathcal{O}_0) = |O_0), \qquad |\mathcal{O}_n) = \frac{|O_n) - \lambda^2 |O_{n-1})}{\sqrt{1-\lambda^4}}\,.
\end{align}
We find that the only nonvanishing Krylov matrix elements are now given by $a_0 = \lambda^2$ and $b_1 = \sqrt{1-\lambda^4}, b_n=1, \forall n>1$, resulting in 
\begin{equation}
\mathcal{U} = \begin{pmatrix}
\lambda^2 & 0 && 0  && 0 & \dots \\
\sqrt{1-\lambda^4} & 0 && 0 && 0 & \dots \\
0 & 1 && 0 && 0 & \dots \\
0 & 0 && 1 && 0 & \dots \\
\vdots & \vdots && \vdots && \vdots & \ddots
\end{pmatrix}\,.
\end{equation}
After a single time step, the new operators generated by the unitary transformation are always linearly independent from all previously generated operators in the Krylov subspace. 
This suggests an interpretation of the set of $|\mathcal{O}_{n>0})$ as structureless, maximally ergodic bath states: The initial operator `leaks' into the bath states during the dynamics, since we can decompose the time-evolved operator as
\begin{align}\label{eq:decompose_DU}
|O_t) = \lambda^{2t}|\mathcal{O}_0) + \sqrt{1-\lambda^4}\sum_{s=1}^t \lambda^{2(t-s)}|\mathcal{O}_s),
\end{align} 
where the bath states do not contribute to the autocorrelation function for $O_0$, since $(O_0|\mathcal{O}_s) = \delta_{s,0}$, and have vanishing autocorrelation functions, $(\mathcal{O}_n |\,\mathcal{U}^t|\mathcal{O}_n) = \delta_{t,0}$.

The expression \eqref{eq:decompose_DU} can be used to directly calculate the Krylov complexity as
\begin{align}
K(t) = t - \frac{\lambda^4}{\lambda^4-1}(\lambda^{4t}-1)\,,
\end{align}
returning linear growth after a time scale $t \propto -1/\log |\lambda|$. While this time scale can be arbitrarily large, in Krylov space the maximal complexity growth is observed after a single step.

These results can also be understood on the level of the spectral function.
Writing $\lambda^2 = e^{-\gamma}$, the spectral function follows as (see also Ref.~\cite{fritzsch_ETH_2021}),
\begin{align}
|f_{O}(\omega)|^2 = \frac{1}{2\pi}\frac{\sinh(\gamma)}{\cosh(\gamma)-\cos(\omega)}.
\end{align}
For large $\gamma$ the autocorrelation function decays rapidly and the spectral function is nearly flat, whereas for small $\gamma$ the autocorrelation functions decays slowly and the spectral function is peaked near $\omega = 0$.

\pwc{
Considering the orthonormal polynomials from Eq.~\eqref{eq:def_orthpoly}, we immediately find that the polynomials associated with the basis transformation to Krylov operators \eqref{eq:DU_Krylovop} are given by
\begin{align}
p_0 = 1,\qquad  p_n(\omega) = e^{i\omega n} \frac{e^{i\omega}-\lambda^2}{\sqrt{1-\lambda^4}}.
\end{align}
}
Furthermore, it is easy to check that $|p_1(\omega)|^{-2} = 2\pi |f_{O}(\omega)|^2$, and hence
\begin{align}
|f_{O}(\omega)|^2 |p_1(\omega)|^2 = \frac{1}{2\pi},
\end{align}
such that $\mathcal{O}_1$ has a flat spectral function. Even though the autocorrelation function for $\mathcal{O}_0$ is nontrivial at all times, the autocorrelation function for $\mathcal{O}_1$ is now a delta function and after a single time step all operators generated by the unitary superoperator are linearly independent from $\mathcal{O}_0$ and $\mathcal{O}_1$. We term this regime to be \emph{maximally ergodic Krylov dynamics}.

\subsection{Maximally ergodic Krylov dynamics}
We can identify three equivalent signifiers of this maximally ergodic regime, where every discrete time step generates an operator that is orthogonal to the previous operators:
\begin{enumerate}
    \item The unitary superoperator becomes purely lower diagonal with $b_n=1$ and $a_n=0$ for $n$ sufficiently large, indicating that $\mathcal{U}|\mathcal{O}_{n}) = |\mathcal{O}_{n+1})$.
    \item The Krylov operators $|\mathcal{O}_n)$ for sufficiently large $n$ have a vanishing autocorrelation function, i.e., all autocorrelations decay instantaneously: $(\mathcal{O}_n|\,\mathcal{U}^t|\mathcal{O}_n)=\delta_{t,0}$. 
    \item For sufficiently large $n$ the orthonormal polynomials satisfy $|f_O(\omega)|^2 |p_n(\omega)|^2 = \textrm{Constant}$, i.e., the spectral functions for $\mathcal{O}_n$ flatten.
\end{enumerate}
We will refer to the Krylov operators in this regime as {\it maximally ergodic}. It is not surprising that such operators can be found in dual-unitary circuits and that they act as attractive fixed points for the Krylov orthonormalization procedure, but we will show in the next section that this behavior is universal in chaotic systems and that the existence of such maximally ergodic operators can serve as a sharp diagnostic for Trotter transitions in integrable systems. The dynamics of these maximally ergodic operators is reminiscent of Bernoulli systems, similarly characterized by instantaneously vanishing correlations \cite{ott_chaos_2002,aravinda_dual-unitary_2021}. Similar exactly solvable Krylov dynamics is observed in Clifford circuits, as shown in Appendix~\ref{app:Clifford}.

While all these statements hold exactly in dual-unitary circuits and Clifford circuits, in more general circuit dynamics we need to allow for some finite error. \pwc{In this case we will take maximally ergodic Krylov dynamics as the regime where the autocorrelation function for the Krylov operators $\mathcal{O}_n$ falls below some small but nonzero threshold, and the matrix elements $a_n$ and $b_n$ in the unitary superoperator reproduce those of the maximally ergodic regime up to this error. 
The precise value of this error only has a small effect on the value of $n$ at which the maximally ergodic regime sets in. In this regime we observe that the autocorrelation functions for Krylov operators $\mathcal{O}_n$ after a single time step decay exponentially with increasing $n$, i.e., $(\mathcal{O}_n|\mathcal{U}|\mathcal{O}_n) \sim e^{-\nu n}$. Reducing the error leads to a logarithmic growth in the value of $n$ at which this regime sets in, but any nonzero error should lead to a finite $n$.}

%{\color{blue} In general circuit dynamics we will take maximally ergodic Krylov dynamics to refer to the regime where the autocorrelation function for the Krylov operators $\mathcal{O}_n$ after a single time step decays exponentially with increasing $n$: $(\mathcal{O}_n|\mathcal{U}|\mathcal{O}_n) \sim e^{-\nu n}$. For any small but nonzero error $(\mathcal{O}_n|\mathcal{U}|\mathcal{O}_n)\sim\epsilon$ we can then find a finite value of $n=n_\epsilon$ and a corresponding Krylov operator $\mathcal{O}_n$, with a support proportional to $n$ and hence scaling logarithmically in the error $n_\epsilon \sim \log(\epsilon)$, for which the autocorrelation function and the matrix elements in the unitary superoperator reproduce the maximally ergodic Krylov dynamics up to this error.} Reducing the error then increases the value of $n$ at which this regime sets in, but any finite error should lead to a finite $n$.

Before continuing, we note that the Krylov dynamics does not just probe properties of the unitary evolution, but also of the initial operator $O_0$. In this respect it is similar to probes of quantum chaos through the spectral function and related measures such as the fidelity susceptibility \cite{sierant_fidelity_2019} and the scaling of the adiabatic gauge potential \cite{kolodrubetz_geometry_2017,pandey_adiabatic_2020}. Even when restricting to local operators, different choices of initial operator can lead to qualitatively different behavior, e.g., integrability-breaking vs. integrability-preserving operators \cite{pandey_adiabatic_2020,brenes_low-frequency_2020}. 
Rather than probing the dynamics in the full operator space, the dependence on the initial operator is made explicit by working within the Krylov subspace particular to this operator \cite{Viswanath1994}. 

\section{Tuning Trotterized Circuits}
\label{sec:trotterized}
In this section we apply the developed framework to a specific class of unitary brickwork circuits, namely, those corresponding to Trotter decompositions of local Hamiltonians~\cite{trotter_product_1959,suzuki_relationship_1976}. Considering the dynamics governed by a local Hamiltonian $H = \sum_{j} H_{j,j+1}$, the unitary evolution operator can be approximated as
\begin{align}
e^{-iHt}\approx \left(U_{\textrm{odd}}(\Delta t) U_{\textrm{even}}(\Delta t) \right)^{t/\Delta t}\,,
\end{align}
with 
\begin{align}
&U_{\textrm{odd}}(\Delta t) = \prod_{j \in 2\mathbb{Z}+1}\left(e^{-i  H_{j,j+1}\Delta t}\right), \\
&U_{\textrm{even}}(\Delta t) = \prod_{j \in 2\mathbb{Z}}\left(e^{-i  H_{j,j+1}\Delta t}\right)\,,
\end{align}
for sufficiently small time step $\Delta t$. This expression reproduces the brickwork structure of Eq.~\eqref{eq:brickwork}, with the local unitary gates given by
\begin{equation}
U_{j,j+1} = \exp[-i H_{j,j+1} \Delta t],
\end{equation}
This decomposition underlies the numerical time-evolving block decimation algorithm~\cite{vidal_efficient_2004,vidal_classical_2007}. However, with the advent of gate-based quantum computation and quantum simulation these circuits can also be realized for larger $\Delta t$, and there has been increasing interest in understanding how the dynamics depends on the choice of Trotter step $\Delta t$~\cite{ishii_heating_2018,sieberer_digital_2019,heyl_quantum_2019,yi_success_2021,chinni_trotter_2022,vernier_integrable_2022,kargi_quantum_2023,zhao_making_2022,keever_classically_2023,kovalsky_self-healing_2023}.

For small $\Delta t$, the Krylov framework for unitary circuits should reproduce the known results for Liouvillian dynamics with $\mathcal{L}(O) = [H,O]$. In this limit the superoperators are related as
\begin{align}\label{eq:L_to_U}
\mathcal{U}(O) &= U^{\dagger}O U \approx e^{i H \Delta t} O e^{-i H \Delta t} \\
&\approx O + i \Delta t [H,O] = (\mathbbm{1}+i \Delta t \mathcal{L})O\,,
\end{align}
up to corrections scaling as $\Delta t^2$. 
As such, we expect the initial Krylov operators $|\mathcal{O}_n)$ to be approximately equal for small $\Delta t$. However, the orthonormalization procedure can be sensitive to small perturbations, and in general the Krylov operators for large $n$ for the unitary superoperator will differ significantly from the Krylov operators for the Liouvillian. As such, we only expect the matrix representations of $\mathcal{U}$ and $\mathcal{L}$ to agree (up to perturbative corrections) for sufficiently small $n$.

In the following, we consider three different classes of Hamiltonians and investigate the corresponding Krylov dynamics.
For chaotic systems we first show how the Liouvillian structure is recovered at small Trotter step and transforms to the behavior of maximally ergodic Krylov dynamics, as observed in dual-unitary circuits, with increasing Trotter step. Next, we turn our attention to interacting integrable circuits with the Trotterized XXZ Hamiltonian as an example. Last we report the results for free-fermionic integrable systems using the XX Hamiltonian as an example.

\subsection{Chaotic Hamiltonians}
\label{subsec:GUE}
\pwc{We consider a minimal model for chaotic Hamiltonians by drawing $H_{i,i+1}$ from the Gaussian unitary ensemble (GUE) and writing $U_{i,i+1}=\exp(-iH_{i,i+1} \Delta t)$.} For different values of $\Delta t$ we numerically perform the orthonormalization procedure using a tensor network approach, representing all operators as matrix product operators (see, e.g., Ref.~\cite{orus_practical_2014}).

\begin{figure}
    \centering
    \includegraphics[width=\columnwidth]{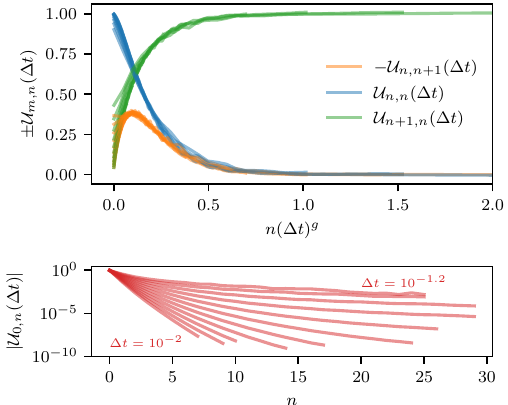}
    \caption{
The unitary superoperator $\mathcal{U}$ [Eq.~\eqref{eq:U_nm}] describing the evolution of the initial operator $|O_0) = \sigma^z_{0}$ when the two-site gate is a Trotterized gate $U_{j,j+1}=\exp(-iH \Delta t)$, where $H$ is drawn from the GUE. \pwc{We represent the defining components of $\mathcal{U}$ in the Krylov basis as in Fig~\ref{fig:tridiagonal_du}.}
%Here $\mathcal{U}$ is given in the Krylov basis, which is generated by orthonormalisation of $\{\mathcal{U}^t|O_0)\}$ with $t\in \mathbb{N}_0$. The unitary superoperator $\mathcal{U}$ is fully described by its tridiagonal matrix elements (top) and its decay away from the diagonal $\mathcal{U}_{n,n+l}=\mathcal{U}_{n,n+1}\mathcal{U}_{0,n+l}/\mathcal{U}_{0,n+1}$ (bottom). 
For small $n (\Delta t)^g$, the matrix elements $\mathcal{U}_{n\pm 1,n}$ coincide with the results obtained for continuous evolution. With increasing $n (\Delta t)^g$, $\mathcal{U}_{n,n}=a_n$ and $\mathcal{U}_{n+1,n}=b_n$ approach $0$ and $1$ respectively (top) and the coefficients $\mathcal{U}_{0,n}=c_n$ decay (bottom), indicating that the unitary superoperator becomes dominated by the tridiagonal elements. The results were obtained for a representative example using a system of size $70$ and TEBD with an MPS bond dimension of 1024 while varying $\Delta t$ from $10^{-2}$ to $10^{-1.2}$ \pwc{ equidistant on a logarithmic scale}. For this range of parameter values $g\approx 1.4$. 
    }
    \label{fig:tridiagonal_chaotic}
\end{figure}

The results for the matrix elements of $\mathcal{U}$ are shown in Fig.~\ref{fig:tridiagonal_chaotic} for different Trotter steps. We here present the matrix elements $\mathcal{U}_{n,n}$ and $\mathcal{U}_{n,n\pm 1}$ for increasing $n$ and different values of $\Delta t$, and we observe a collapse of the results for different Trotter steps if we plot the matrix elements as a function of $n (\Delta t)^{g}$, with $g\approx 1.4$. Before discussing the origin of this scaling collapse, we first observe that two different regimes with an intermediate crossover can be distinguished. 

\emph{Hamiltonian dynamics}. 
The first regime appears at small values of $n (\Delta t)^{g}$ and reproduces the expected results from Hamiltonian evolution: up to $n \propto 1/\Delta t$, the matrix elements of $\mathcal{U}$ reproduce those of the Liouvillian $\mathcal{L}$ as discussed in Ref.~\cite{parker_universal_2019}. To be more precise, we find that 
\begin{align}
&|a_n -1| \propto n^{1.4} \Delta t^2\,, \\
&|\tilde{b}_n + b_n/\Delta t| \propto (n \Delta t)^2\,,
\end{align}
for $n \Delta t \lesssim 1$. 
Note that in this regime the unitary superoperator is close to being tridiagonal. The diagonal elements being close to one reflects the appearance of the identity in Eq.~\eqref{eq:L_to_U}, whereas the off-diagonal elements $b_n$ reproduce the off-diagonal elements $\tilde{b}_n$ from the Liouvillian dynamics up to a rescaling by $\Delta t$. In this regime, the operator dynamics of the Trotterized circuit can hence be expected to remain close to the operator dynamics of the original Hamiltonian.

However, this regime cannot persist for arbitrarily large values of $n$. In chaotic systems the off-diagonal elements $\tilde{b}_n$ in the Liouvillian are expected to grow with the maximal possible rate as $n$ increases, corresponding to growth in the off-diagonal element of $\mathcal{U}$ since $b_n \approx -\tilde{b}_n \Delta t$. However, unitarity restricts the matrix elements of $\mathcal{U}$ to satisfy $|\mathcal{U}_{mn}| \leq 1$, such that the growth of $b_n$ is necessarily bounded.

\emph{Maximally ergodic dynamics}. 
The second regime appears at larger values of $n $. Here $b_n = \mathcal{\mathcal{U}}_{n+1,n}\approx 1$ and all other matrix elements $\mathcal{U}_{m,n} \approx 0$ for $m\neq n+1$, with the errors decreasing exponentially with increasing $n$ (\pwc{see also Appendix~\ref{app:convergence_large_n}}). This structure of the Krylov matrix is identical to the structure previously observed in dual-unitary circuits in Eq.~\eqref{eq:DU_Krylov}, indicating that in each step the Heisenberg evolution with $U$ creates a new operator that is linearly independent from all previously generated operators in the Krylov subspace. In this regime the operator dynamics in the unitary circuit is qualitatively different from that of the Hamiltonian system. As one example, we can again consider the autocorrelation function for a Krylov operator $|\mathcal{O}_n)$. In the circuit case, the autocorrelation identically vanishes after a single discrete time step (up to exponentially small corrections), same as in dual-unitary circuits, which is prohibited in the Hamiltonian case. Here we can also contrast the long-time behavior of the autocorrelation function, since in the Hamiltonian case the autocorrelation function is expected to reach a nonzero thermal value constrained by conservation of energy, again to be contrasted with the vanishing autocorrelation function in the unitary circuit. 

\begin{figure}
    \centering
    \includegraphics[width=1.0\columnwidth]{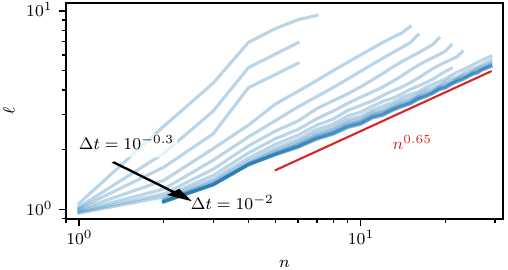}
    \caption{Spatial extent $\ell$ of the Krylov operators $|\mathcal{O}_n)$ with increasing $n$. The extent is measured by fitting the error function to the OTOC calculated for each $|\mathcal{O}_n)$ for different $\Delta t$. Here $\ell$ is estimated by the point where the error function takes half its maximum value. For small $\Delta t$, the extent of the operator grows roughly as $\ell \propto n^{0.65}$. The same $\mathcal{U}$ and simulation setting was used as in Fig.~\ref{fig:tridiagonal_chaotic}. For details see Appendix~\ref{app:fit_error_function}. 
    \label{fig:operator_extent}}
\end{figure}

\emph{Crossover regime.}
The remaining question is about the origin of the transition scale and the collapse observed when recscaling $n$ with $\Delta t^{g}$. As it turns out, this scale is very closely related to the size of the operator and operator spreading. \pwc{ Expanding $\mathcal{U} |\mathcal{O}_n)$ in $\Delta t$, the first order term is of magnitude $\ell \Delta t$, where $\ell$ is the number of sites $\mathcal{O}_n$ has support on. Hence,} we expect a transition in the behavior of the circuit once an operator $|\mathcal{O}_n)$ acts non-trivially on around $\ell\sim 1/\Delta t$ sites. While the full support of both $|O_n)$ and $|\mathcal{O}_n)$ grows linearly in $n$, the bulk of the operator typically acts on a smaller number of sites, as quantified in, e.g., out-of-time-order correlators (OTOCs) and operator spreading~\cite{nahum_operator_2018,von_keyserlingk_operator_2018,khemani_operator_2018}. While the support of $|O_n)$ is by now well understood through studies in, e.g., random unitary circuits, these results do not directly translate to the support of $|\mathcal{O}_n)$. We quantify the operator growth in the usual way, considering 

\begin{align}
\mathrm{OTOC}(n,j) = \mathrm{Tr}\left[ \mathcal{O}_n^\dagger\sigma^\gamma_j \mathcal{O}_n \sigma^\gamma_j \right] /\mathcal{D}\,,
\end{align}
with $\gamma \in \{x,y,z\}$.
We observe that this profile resembles an error function, as also expected for $|O_n)$, and define $\ell$ as the value of $n$ where the error function equals $1/2$. The results are shown in Fig.~\ref{fig:operator_extent}. For the considered range of $\Delta t$, we have that $\ell \propto n^{0.65}$. In combination with ${\ell} \sim 1/\Delta t$ this roughly yields the observed transition on a scale $n \sim \Delta t^{-1.5}$. However, we note that this exponent is not universal since it crucially depends on $\Delta t$, which is apparent when considering different limits. 

For large $\Delta t$ we expect the operator spreading for $|\mathcal{O}_n)$ and $|O_n)$ to coincide and exhibit the biased diffusive growth previously observed~\cite{von_keyserlingk_operator_2018,khemani_operator_2018,nahum_operator_2018}, corresponding to an operator growth ${\ell} \propto n$. For very small $\Delta t$, we can approximate the unitary superoperator by the Liouvillian. In this limit the operator growth is subballistic, as most of the local operators in the Hamiltonian act on the bulk of the operator $|\mathcal{O}_n)$. We hence expect the operator growth to increase from subballistic growth in the limit $\Delta t\rightarrow 0$ to ballistic growth for large $\Delta t$, with an intermediate superdiffusive but nonuniversal regime corresponding to the scaling observed in Fig.~\ref{fig:tridiagonal_chaotic}.
\begin{figure*}[ht]
	\begin{center}
    \centering
    \includegraphics{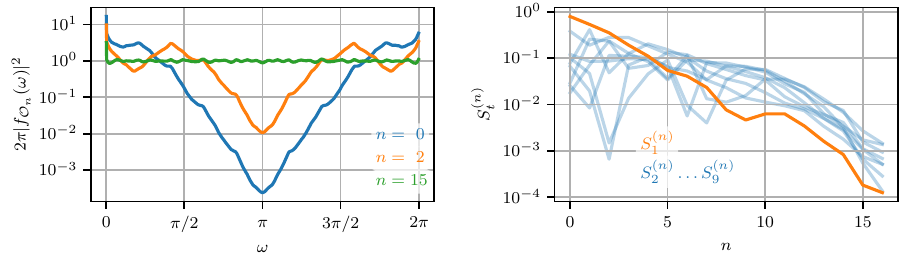}    
    \caption{Spectral function $|f_O(\omega)|^2 = \sum_{p,q} |\langle p|O|q \rangle|^2\, \delta(\theta_q-\theta_p-\omega)/\mathcal{D}$ (left) and its Fourier transform (right), i.e., the autocorrelation functions or Fourier modes $S^{(n)}_t = \mathrm{Tr}[\mathcal{O}_n^\dagger U^{\dagger t} \mathcal{O}_n U^t]/\mathcal{D}$, for the Krylov operators $\mathcal{O}_n$ for the GUE system defined in Eq.~\eqref{eq:spectral_function_for_operator}. For $n\rightarrow \infty$, the spectral function approaches a constant. This approach is quantified through the Fourier modes of the spectral function of $\mathcal{O}_n$, where all Fourier modes $S_t^{(n)}$ of the spectral function appear to decay exponentially with $n$. This decay implies the exponential decay with $n$ of the autocorrelation function of $\mathcal{O}_n$ for all times. The first Fourier mode $S_1^{(n)}$ is highlighted in orange as it is directly related to the unitary superoperator via $S_1^{(n)}=\mathcal{U}_{nn}=a_n$ [Eq.~\eqref{eq:relat_spec_prop_and_modes}]. These results were obtained by exact diagonalization of a system of size $14$ with open boundary conditions using the same circuit structure and Hamiltonian parameters as in Fig.~\ref{fig:tridiagonal_chaotic}. Here we choose $\Delta t = 10^{-0.8}$, but the results are qualitatively similar for different $\Delta t$. To mitigate finite size effects, $|f_{\mathcal{O}_n}|^2$ is averaged over $\Delta \omega = 2\pi/1000$.}
    \label{fig:spectral_GUE}
	\end{center}
 \vspace{-\baselineskip}
\end{figure*}

\emph{Spectral function.} \pwc{ The results from the previous section can be understood on the level of the spectral function, which 
%also allows us to make exact statements about
yields a sufficient condition for the existence of the maximally ergodic regime as a stable attractor for the orthonormalization procedure.} On the level of the spectral function, maximally chaotic dynamics corresponds to the spectral function of the Krylov operators being a flat function. These spectral functions are plotted in the left panel of Fig.~\ref{fig:spectral_GUE} for different values of $n$, and we indeed observe that the spectral function flattens with increasing $n$.

Following Eq.~\eqref{eq:spectral_On}, these spectral functions are related to the spectral function of the initial operator as 
\begin{align}
|f_{\mathcal{O}_n}(\omega)|^2 = |f_O(\omega)|^2 |p_n(\omega)|^2\,,
\end{align}
where $p_n(\omega)$ are the polynomials representing the Krylov operators, which are orthonormal polynomials with $|f_O(\omega)|^2$ as weight functions. Stating that the Krylov operators flow to the maximally chaotic regime is \pwc{equivalent to the convergence of the polynomials stated in Eq.~\eqref{eq:convergence_pn_0}.}
%saying that
%%%
%\begin{align}
%\lim_{n \to \infty}|f_O(\omega)|^2 |p_n(\omega)|^2 = \frac{1}{2\pi},
%\end{align}
%%%
%or, equivalently,
%%%
%\begin{align}\label{eq:convergence_pn}
%\lim_{n \to \infty}|p_n(\omega)|^{-2} %= \frac{1}{2\pi}|f_O(\omega)|^{2}\,.
%\end{align}
%%%
As mentioned in Sec.~\ref{subsec:spectral}, such a (uniform) convergence has in fact already been established in the theory of orthonormal polynomials on the unit circle provided the weight function/spectral function satisfies certain properties ~\cite{szego_orthogonal_1975,nevai_geza_1986}. However, this convergence crucially depends on the original spectral function $\log(|f_O(\omega)|^2)$ being Lebesgue integrable. This Lebesgue integrability is guaranteed if this spectral function is nonvanishing on the entire interval $[0,2\pi)$, in which case the logarithm of the spectral function is well-defined and \pwc{finite, yielding a sufficient condition for the existence of the maximally chaotic regime. This condition is expected to be fulfilled on physical grounds for chaotic systems: A vanishing spectral function would imply the existence of either selection rules or a spectrum that is not dense on the unit circle, both prohibited in chaotic dynamics~\cite{dalessio_long-time_2014,lazarides_equilibrium_2014}.}
More specifically, these orthonormal polynomials have the property that the first $n$ Fourier modes of $|p_n(\omega)|^{-2}$ agree with the first $n$ Fourier modes of $|f_O(\omega)|^{2}$. For $n\to \infty$ all moments agree, indicating that Eq.~\eqref{eq:convergence_pn_0} indeed applies (up to pathological exceptions, see also Section~\ref{sec:noninteracting}). 

The diagonal of the unitary superoperator, $a_n = \mathcal{U}_{nn}$, is directly related to the spectral properties via
\begin{align}\label{eq:relat_spec_prop_and_modes}
    \mathcal{U}_{nn} = (\mathcal{O}_n|\,\mathcal{U}|\mathcal{O}_n) = \int_0^{2\pi} \!\mathrm{d}\omega \, e^{i\omega} |f_{\mathcal{O}_n}(\omega)|^2 \equiv S_1^{(n)}\,.
\end{align}
This expression relates the exponential decay of the diagonal of the unitary superoperator to the exponential decay of the first Fourier component $S_1^{(n)}$ of the spectral function of $\mathcal{O}_n$. 
We can more generally consider the decay of arbitrary Fourier modes
\begin{align}
 S_t^{(n)} = \int^{2\pi}_0\!\mathrm{d}\omega\,|f_{\mathcal{O}_n}(\omega)|^2 e^{i\omega t}%\,,
 \end{align}
with $t \in \mathbb{N}_0$ as illustrated in the right panel of Fig.~\ref{fig:spectral_GUE}. All Fourier modes $S_t^{(n)}$ appear to decay similarly with increasing $n$, as also expected by the uniform convergence in Eq.~\eqref{eq:convergence_pn_0}. Hence, we expect in general a uniform convergence of the spectral function $|f_{\mathcal{O}_n}|^2$ towards $1/2\pi$, agreeing with the convergence of the unitary superoperator towards its maximally Krylov ergodic form. 

We note that this convergence is particular to orthonormal polynomials on the unit circle and hence to unitary circuit dynamics. Conversely, the universal operator growth hypothesis for Hamiltonian dynamics translates to statements about the large-frequency asymptotics of the spectral function \cite{parker_universal_2019}.

\subsection{Interacting integrable Hamiltonians}
\begin{figure}
    \centering
    \includegraphics[width=\columnwidth]{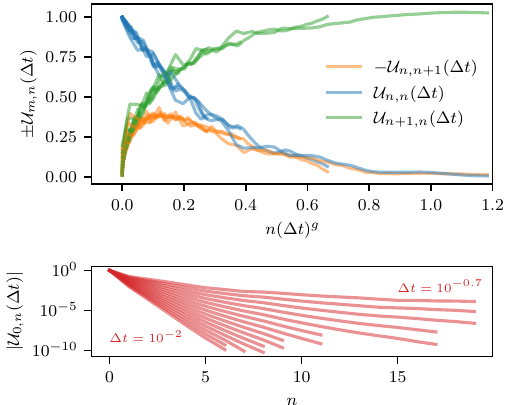}
    \caption{
The unitary superoperator $\mathcal{U}$ [Eq.~\eqref{eq:U_nm}] describing the evolution of the initial operator $|O_0) = \sigma^z_{0}$ when the two-site gate is a Trotterized gate $U_{j,j+1}=\exp(-iH \Delta t)$, where $H$ is the XXZ Hamiltonian with $\Delta=3$. \pwc{We represent the defining components of $\mathcal{U}$ in the Krylov basis as in Fig~\ref{fig:tridiagonal_du}.}
%Here $\mathcal{U}$ is given in the Krylov basis.  The unitary superoperator $\mathcal{U}$ is fully described by its tridiagonal matrix elements (top) and its decay away from the diagonal $\mathcal{U}_{n,n+l}=\mathcal{U}_{n,n+1}\mathcal{U}_{0,n+l}/\mathcal{U}_{0,n+1}$ (bottom). 
The results are qualitatively similar to the GUE case, see Fig.~\ref{fig:tridiagonal_chaotic}. For small $n (\Delta t)^g$, the matrix elements $\mathcal{U}_{n\pm 1,n}$ coincide with the results obtained for continuous evolution. With increasing $n (\Delta t)^g$, $\mathcal{U}_{n,n}=a_n$ and $\mathcal{U}_{n+1,n}=b_n$ approach $0$ and $1$ respectively (top) and the coefficients $\mathcal{U}_{0,n}=c_n$ decay (bottom), indicating that the unitary superoperator becomes dominated by the tridiagonal elements. The results were obtained using a system of size $70$ and TEBD with an MPS bond dimension of 1024 while varying $\Delta t$ from $10^{-2}$ to $10^{-0.7}$. For this range of parameter values $g\approx 2.1$. 
}
    \label{fig:tridiagonal_xxz}
\end{figure}

Let us now consider the integrable XXZ Hamiltonian, 
\begin{align}
H_{j,j+1}=\sigma_j^x \sigma_{j+1}^x+\sigma_{j}^y \sigma_{j+1}^y+\Delta \sigma_{j}^z \sigma_{j+1}^z.
\end{align}
For small $\Delta t$ the matrix elements converge towards those of the Liouvillian for the Hamiltonian case, where $\mathcal{L}_{n+1,n}\propto \sqrt{n}$, see Fig.~\ref{fig:tridiagonal_xxz}. However, for larger $n$ the qualitative difference between the integrable and chaotic dynamics disappears and we recover the maximally ergodic form with only ones on the lower diagonal and zeros everywhere else, on the same scales as for the chaotic model. This convergence is not unexpected, since the arguments presented in the previous subsection for the convergence of the spectral function directly carry over to the integrable model, as the spectral function for generic operators is expected to be nonvanishing on the entire interval $[0,2\pi]$ (which is not guaranteed for noninteracting systems).

The difference in Krylov dynamics for chaotic and interacting integrable remains a topic of active research.
Let us briefly discuss some recent results for the Krylov dynamics for interacting integrable systems under Hamiltonian evolution (see Refs.~\cite{rabinovici_krylov_2022,rabinovici_krylovlocalization_2022}) and relate these to the observed phenomenology. For finite systems, the Krylov complexity saturates at Heisenberg time scales $t_H \propto \mathcal{D}$.  In chaotic systems, this saturation value was argued to be approximately $\mathcal{D}^2/2$, whereas in interacting integrable systems the Krylov complexity saturates at a smaller value (for which no analytic prediction exists). These saturation values can be linked to localization in Krylov space: delocalization in Krylov space results in the chaotic saturation value, whereas a certain degree of localization in Krylov space leads to a saturation value below the chaotic one. Crucially, within interacting integrable systems this localization is induced by disorder in the Lanczos coefficients, where the Lanczos coefficients in interacting integrable system are more disordered than these in chaotic systems. The same behavior can be observed in the unitary superoperator (Fig.~\ref{fig:tridiagonal_xxz}), where the coefficients are noisier compared to the GUE results (Fig.~\ref{fig:tridiagonal_chaotic}). 

An additional difference between chaotic and interacting integrable can be observed here: while both chaotic and interacting integrable systems converge to the maximally ergodic regime, the number of discrete time steps required to reach this regime appears to be (parametrically) larger in interacting integrable systems. 
As discussed in the GUE example, the unitary dynamics can be expected to reproduce the Hamiltonian dynamics for sufficiently small $n$ provided the matrix elements of $\mathcal{U}$ reproduce those of the Liouvillian $\mathcal{L}$. For chaotic systems, we already observed that this regime holds up to $n \sim 1/\Delta t$. 
Conversely, in interacting integrable systems this regime holds up to $n \sim 1/\Delta t^{1.3}$ (with an exponent that we do not expect to be universal) using
\begin{align}
&|a_n -1| \propto n^{1.5} \Delta t^2\,, \\
&|\tilde{b}_n + b_n/\Delta t| \propto n^{1.5}\Delta t^2\,
\end{align}
as numerically shown in Appendix~\ref{sec:gue_convergence_hermtian}. As such, it takes parametrically longer to reach the maximally ergodic regime in interacting integrable circuits as compared to the chaotic case. The larger values of $n$ required to reach this regime can also be understood from the slower growth of $b_n$ for small $n$ as compared to chaotic systems, since the maximally ergodic regime is characterized by a saturation of $b_n$. However, the difference between chaotic and interacting integrable Krylov dynamics remains largely quantitative rather than qualitative (as for Hamiltonian dynamics).

\subsection{Noninteracting integrable Hamiltonians}
We now consider integrable systems that can be mapped to free fermions. We will first restrict ourselves to numerical results, and analytical arguments will be presented in Sec.~\ref{sec:noninteracting}. As a representative example, we focus on
\begin{align}
H_{j,j+1}=\sigma_j^x \sigma_{j+1}^x+\sigma_{j}^y \sigma_{j+1}^y\,,
\end{align}
and consider a local initial operator $\sigma^z_0$. 

\begin{figure}
    \centering
    \includegraphics[width=\columnwidth]{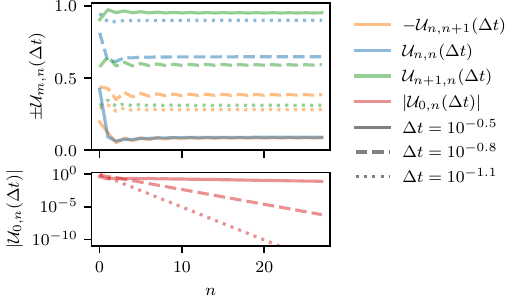}
    \caption{
    The unitary superoperator $\mathcal{U}$ [Eq.~\eqref{eq:U_nm}] describing the evolution of the initial operator $|O_0) = \sigma^z_{0}$ when the two-site gate is a Trotterized gate $U_{j,j+1}=\exp(-iH \Delta t)$, where $H$ is the XX Hamiltonian. \pwc{We represent the defining components of $\mathcal{U}$ in the Krylov basis as in Fig~\ref{fig:tridiagonal_du}. }%, which is generated by orthonormalisation of $\{\mathcal{U}^t|O_0)\}$ with $t\in \mathbb{N}_0$. 
%The unitary superoperator $\mathcal{U}$ is fully described by its tridiagonal matrix elements (top) and its exponential decay away from the diagonal $\mathcal{U}_{n,n+l}=\mathcal{U}_{n,n+1}\mathcal{U}_{0,n+l}/\mathcal{U}_{0,n+1}$ (bottom). 
After an initial transient behavior the diagonals approach constant values; the dependence of those values on $\Delta t$ is shown in Fig.~\ref{fig:tridiagonal_free_ferm}. These results were obtained using the analytical expressions [Eq.~\eqref{eq:analyt_exp_free_ferm}].
    }
    \label{fig:tridiagonal_xx}
\end{figure}

\begin{figure}
    \centering
    \includegraphics[width=\columnwidth]{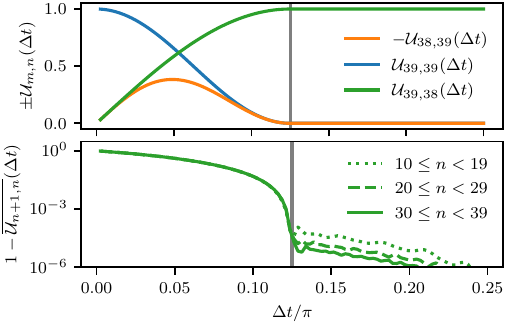}
    \caption{Constant values approached by the tridiagonal part of $\mathcal{U}$ for large $n$, see Fig.~\ref{fig:tridiagonal_xx}, as function of $\Delta t$. A transition is visible at $\Delta t = \pi/8$, where the value of the lower diagonal $\overline{U}_{n+1,n}$ approaches $1$ nonanalytically for $n\rightarrow \infty$ (lower panel). The results were obtained by numerical diagonalization of a system of size $200$ with open boundary conditions. \pwc{Data points for the lower diagonal are averaged over the ten indicated values of $n$ for ease of visualization.}}
    \label{fig:tridiagonal_free_ferm}
\end{figure}

For these circuits we again find the diagonal part of the Krylov matrix $\mathcal{L}$ to be dominant, see Fig.~\ref{fig:tridiagonal_xx}. However, the noticeable difference in comparison to the chaotic and interacting integrable circuits is the large $n$ limit. We observe that for large $n$, $b_n = \mathcal{U}_{n+1,n}$ approaches a constant value which is lower than $1$ for $\Delta t < \pi/8$ and $1$ for $\Delta t \geq \pi/8$, see Fig.~\ref{fig:tridiagonal_free_ferm}. As such, for free-fermionic integrable models the existence of the maximally chaotic regime crucially depends on the choice of Trotter step: for a small Trotter step $\mathcal{U}$ never approaches the maximally ergodic Krylov dynamics. 
This transition can be understood analytically, as will be made apparent in the next section.

\section{Trotter transition in noninteracting circuits}
\label{sec:noninteracting}
We here detail the dynamics of the integrable non-interacting system, presenting exact results for the autocorrelation functions and corresponding spectral functions, and relate these to the nonanalytic Trotter transitions observed in the previous section. 
A defining feature of noninteracting models is that their dynamics can be decomposed in the dynamics of decoupled single-particle sectors. Trotter decompositions of noninteracting models have the property that the corresponding unitary circuit remains noninteracting since such a decoupling in single-particle sectors remains possible. This can be understood as a specific case of a Floquet drive of noninteracting models that does not couple different particle sectors, preserving the noninteracting structure and leading to a noninteracting Floquet Hamiltonian~\cite{gritsev_integrable_2017}. These Trotter transitions can also be seen as transitions in the Floquet Hamiltonian as the driving frequency, as set by $\Delta t$, is varied.

The two-site unitary gate underlying this circuit is given by
\begin{align}
U_{12}= \exp\left[-i \Delta t (\sigma_1^x \sigma_2^x + \sigma_1^y \sigma_2^y)\right]\,.
\end{align}  
We first note that at $\Delta t = \pi/4$ this gate reduces to the iSWAP gate, i.e.,
\begin{align}
U = 
\begin{pmatrix}
1 & 0 & 0 & 0 \\
0 & 0 & i & 0 \\
0 & i & 0 & 0 \\
0 & 0 & 0 & 1
\end{pmatrix}\,,
\end{align}
which is a particular case of a dual-unitary gate.

The dynamics of the full circuit can be exactly solved through a Jordan-Wigner transformation followed by a decomposition in Fourier modes with fixed momentum. A detailed derivation is given in Appendix~\ref{app:detailed_der_integrable_system}, and we here only outline the final results. The Jordan-Wigner transformation maps the spin operators to fermionic operators, and we consider fermionic single-particle operators with fixed momentum acting on even and odd sites respectively,
\begin{align}
&c_{k+} = \frac{1}{\sqrt{N}}\sum_{j=0}^{N-1} e^{ikj} c_{2j}, \\
&c_{k-} = \frac{1}{\sqrt{N}}\sum_{j=0}^{N-1} e^{ikj} c_{2j+1},
\end{align}
where $c_{j}$ acts on site $j$ and satisfies the fermionic anticommutation relation $\{c_i^\dagger,c_j\}=\delta_{ij}$. We consider a system consisting of $2N$ sites, and $k$ can take the values $2 \pi n/N, n=0,1, \dots , N-1$. Denoting $|c_{k\pm})$ as $|k_\pm)$, we find that the unitary superoperator acting on fermionic annihilation operators can be decomposed as
\begin{align}\label{eq:U_k}
\mathcal{U} = \sum_{k}
\begin{pmatrix}
|k_+) & |k_-)
\end{pmatrix}
\mathcal{U}_k
\begin{pmatrix}
(k_+| \\ (k_-|
\end{pmatrix},
\end{align}
where $\mathcal{U}_k$ are $2 \times 2$ matrices parametrized as
\begin{align}\label{eq:analyt_exp_free_ferm}
\mathcal{U}_k &= \mathbbm{1}-\sin(2 \Delta t) \nonumber\\
&\times \begin{pmatrix}
\sin(2 \Delta t)(1+e^{ik}) & i \cos(2 \Delta  t)(1+e^{ik})  \\
 i \cos(2 \Delta t) (1+e^{-ik})&\sin(2 \Delta  t) (1+e^{-ik})
\end{pmatrix}\,.
\end{align}  
For fermionic creation operators the Hermitian conjugate of this matrix should be used.

Equation~\eqref{eq:U_k} can be directly diagonalized to obtain two complex-conjugate eigenvalues $e^{\pm i \omega_k}$, where the dispersion relation follows as
\begin{align}
    \cos(\omega_k) =1 - 2 \sin^2(2\Delta t)\cos^2(k/2).
\end{align}
In the limit $\Delta t \to 0$, we recover the dispersion relation for the original Hamiltonian dynamics, with
\begin{align}
\omega_k = 4 \Delta t \cos(k/2)\,,
\end{align}
up to corrections scaling as $\Delta t^2$. At $\Delta t = \pi/4$ the corresponding brickwork circuit is dual unitary, and we find that
\begin{align}
\omega_k = k \pm \pi\,.
\end{align}
Here the linear dispersion indicates that all excitations spread ballistically with a fixed velocity $\pm 1$, as readily apparent for iSWAP gates and more generally expected for integrable dual-unitary gates, where conservation laws correspond to solitons with maximal velocity $\pm 1$ \cite{bertini_solitons_2020}. The appearance of the factor $\pi$ indicates a sign change after every time step, which is a consequence of applying the iSWAP gate twice.

The Krylov dynamics is fully determined by the autocorrelation function. For, e.g., $O_0 = \sigma_0^z$, the autocorrelation follows as
\begin{align}\label{eq:nonint_autocorr_S0z}
\frac{1}{2}\mathrm{Tr}\left[\sigma_0^z(t)\sigma_0^z\right] = \left|\mathrm{Tr}\left[c_0^{\dagger}(t)c_0\right]\right|^2\,,
\end{align}
where we used $\sigma^z_0 = 1-2c_0^\dagger c_0$ and the fermionic commutation relations. We can use the decomposition in eigenmodes to obtain (see Appendix \ref{app:detailed_der_integrable_system}) 
\begin{align}\label{eq:corr_c0}
\mathrm{Tr}\left[c_0^{\dagger}(t)c_0\right] = \frac{1}{2\pi} \int_0^{2\pi} \!\mathrm{d}k \cos(\omega_k t),
\end{align}
where we have taken the thermodynamic limit of an infinite system size, $N \to \infty$, in the previous expressions.

\begin{figure}[h]
    \centering
    \includegraphics{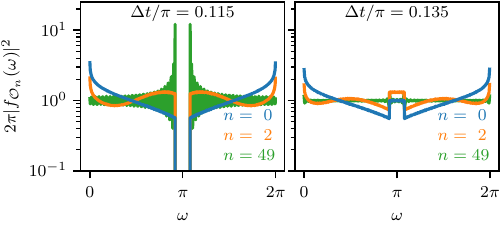}

\quad

    \includegraphics{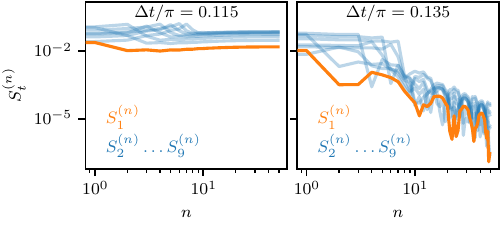}
    \caption{Spectral function $|f_O(\omega)|^2 = \sum_{p,q} |\langle p|O|q \rangle|^2\, \delta(\theta_q-\theta_p-\omega)/\mathcal{D}$ (top) and its Fourier transform (bottom), i.e., the autocorrelation function or Fourier modes $S^{(n)}_t = \mathrm{Tr}[\mathcal{O}_n^\dagger U^{\dagger t} \mathcal{O}_n U^t]/\mathcal{D}$, for the Krylov operators $\mathcal{O}_n$ for the non-interacting integrable system based on the Trotterized XX model. Here $|O_0) = \sigma^z_0$. For $\Delta t < \pi/8$, the spectral function is gapped and the autocorrelation, i.e., Fourier modes $S_t^{(n)}$, does not decay with $n$, while the gap is closed for $\Delta t \geq \pi/8$ yielding a decaying autocorrelation. 
    The results were obtained by numerical diagonalization of a system of size $200$ with open boundary conditions. To mitigate finite size effects, $|f_{\mathcal{O}_n}|^2$ is averaged over $\Delta \omega = 2 \pi /1000$.
    \label{fig:spectral_XX}}
\end{figure}

From this expression it follows that the spin autocorrelation function \eqref{eq:nonint_autocorr_S0z} will contain oscillations with frequencies $\omega_{k}\pm\omega_{k'}$. For small $\Delta t$ these are bounded by $2 \omega_{k=0}$, where the dispersion relation takes a particularly simple form,
\begin{align}
\omega_{k=0} = 4 \Delta t\,.
\end{align}
As such, the spectral function has support in $[-2 \times 4 \Delta t, 2 \times 4 \Delta t]$, as illustrated in Fig.~\ref{fig:spectral_XX}. We hence find that the spectral function is gapped around the $\pi$-mode where $\omega_k=\pi$ for $0 \leq \Delta t < \pi/8$, the gap at the $\pi$-mode closes at $\Delta t = \pi/8$, and then the system remains gapless for $\pi/8 \leq \Delta t \leq 3\pi/8$. In the gapped phase the spectral function has a gap, whereas the spectral function is fully supported on $[0,2\pi]$ in the gapless phase. The unitary gate is periodic in $\Delta t$, so a further increase of $\Delta t$ would lead to alternating gapped and gapless phases.

The spectral function for $S_0^z$ can be analytically obtained from Eq.~\eqref{eq:corr_c0}. Focusing on the case $0 \leq \Delta t < \pi/8$ and performing a change of integration variable in Eq.~\eqref{eq:corr_c0} from $k$ to $\omega_k$, we find
\begin{align}
    \mathrm{Tr}\left[c_0^{\dagger}(t)c_0\right] = \int_{-4 \Delta t}^{4 \Delta t}\!\mathrm{d}\omega_k \frac{\cos(\omega_k/2)\cos(\omega_k t)}{\sqrt{\cos^2(\omega_k/2)-\cos^2(2\Delta t)}},
\end{align}
such that the spectral function for $c_0^{\dagger}$ can be read off as
\begin{align}
    |f_{c_0}(\omega)|^2 =\frac{\cos(\omega/2)}{\sqrt{\cos^2(\omega/2)-\cos^2(2\Delta t)}}\,.
\end{align}
for $|\omega|<4 \Delta t$ and zero otherwise. The spectral function for the spin operator $\sigma_0^z$ then follows from the convolution theorem as
\begin{align}
     |f_O(\omega)|^2 = \frac{1}{2\pi} \int_{-4 \Delta t}^{4 \Delta t} \!\mathrm{d}\mu\, |f_{c_0}(\mu)|^2 |f_{c_0}(\omega-\mu)|^2\,.
\end{align}
This expression corresponds to the spectral function from Fig.~\ref{fig:spectral_XX} in the gapped phase.

The action of the Krylov orthonormalization procedure can again be understood as a flattening of the spectral function, as illustrated in Fig.~\ref{fig:spectral_XX}. However, if $|f_O(\omega)|^2=0$ for some values of $\omega$, then at these values we also have that $|f_{\mathcal{O}_n}(\omega)|^2 = |f_O(\omega)|^2|p_n(\omega)|^2=0$, i.e., the spectral function of $|\mathcal{O}_n)$ also vanishes. As such, for a gapless phase and a spectral function that is fully supported on $[0,2\pi]$ we can repeat the arguments from Sect.~\ref{subsec:GUE} to argue that the Krylov operators approach the maximally ergodic form where $b_n=1$ and $(\mathcal{O}_n|\mathcal{U}|\mathcal{O}_n) = 0$ for large $n$. However, this is impossible in the gapped phase, since the existence of such maximally ergodic operators requires a constant spectral function, which is precluded by the gap in $|f_{\mathcal{O}_n}(\omega)|^2$. 
The nonanalytic transition in the Krylov dynamics can hence be directly related to the presence of a gap at the $\pi$-mode in the spectral function. 

\begin{figure}
    \centering
    \includegraphics[width=\columnwidth]{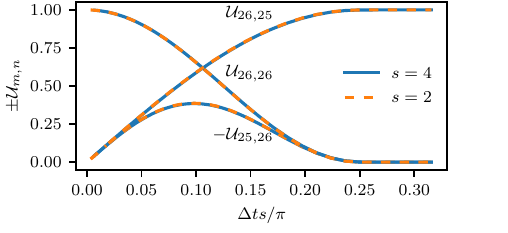}
    \caption{Tridiagonal matrix elements of the unitary superoperator $\mathcal{U}$ for the noninteracting dynamics and initial operators with varying fermionic sizes $s$. The results were obtained using the autocorrelation function $|\mathrm{Tr}[c_0^\dagger(t)c_0]|^s$, Eq.~\eqref{eq:corr_function_s}. Rescaling $\Delta t$ with $s$ yields a collapse. The results were obtained using a system of size $2000$ with periodic boundary conditions and an increased numerical precision of $100$ digits.
    \label{fig:critical_behaviour_xx}}
\end{figure}

The previous analysis depended on the fact that the initial spin operator could be recast as a fermion bilinear. 
This analysis can be directly extended to operators of increasing size. If an initial operator $O_0$ can be decomposed in $s$ fermionic operators, the frequencies appearing in the Fourier transform of its autocorrelation will be of the form $\pm \omega_{k_1} \pm \omega_{k_2} \dots \pm \omega_{k_{s}}$.  
Hence, if the single-particle spectrum $\omega_k$ has support $[-4\Delta t, 4\Delta t]$, the spectrum of the $s$-particle operator has extend $[-4s\Delta t, 4s\Delta t]$. Indeed, we find for general operators that the transition appears at $\Delta t = \pi/(4s)$, as illustrated in Fig.~\ref{fig:critical_behaviour_xx} for an operator with autocorrelation function
\begin{align}\label{eq:corr_function_s}
    \left|\mathrm{Tr}\left[c_0^\dagger(t)c_0\right]\right|^s.
\end{align}
This autocorrelation function corresponds to that of operators $|O_0) = c_{0}$ ($s=1$), $|O_0) = \sigma^z_{0}$ ($s=2$), and more generally $\sigma^z_{0}\sigma^z_{m} \dots \sigma^z_{(m-2) s/2}$ (arbitrary $s$) for spatially separated $\sigma^z$ operators with a sufficiently large spacing $m$ such that the autocorrelation function factorizes.
This result highlights the dependence of the chaotic nature of the operator dynamics on the initial operator. The larger the operator, the larger the regime where it is maximally Krylov ergodic. A similar effect was observed in Hamiltonian evolution, where the Krylov dynamics of a nonlocal operator in a noninteracting systems resembles the Krylov dynamics of an operator in an interacting systems~\cite{parker_universal_2019}.

\pwc{To conclude this section, we briefly contrast the Trotter transition observed in this work with the Trotter transitions discussed in Refs.~\cite{sieberer_digital_2019,kargi_quantum_2023}. These works consider integrable Hamiltonians for which the Trotterization breaks integrability for all nonzero Trotter steps. As the Trotter step is increased the integrability-breaking effects become more important and lead to a sharp transition to chaotic dynamics. In this work, conversely, we consider Trotterizations that do not break integrability, and these transitions happen while the full circuit remains (noninteracting) integrable. In this sense these transitions are comparable to heating transitions in periodically driven noninteracting system as the driving frequency is varied, which similarly happen without breaking integrability~\cite{prosen_nonequilibrium_2011,sen_entanglement_2016,tapias_arze_out--equilibrium_2020}.}

\section{Krylov complexity}
\label{sec:complexity}
\begin{figure*}
\includegraphics{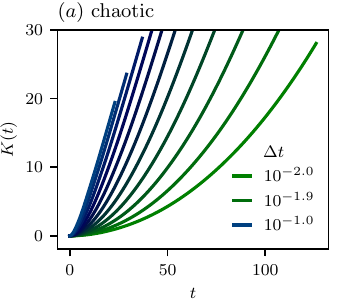}
\includegraphics{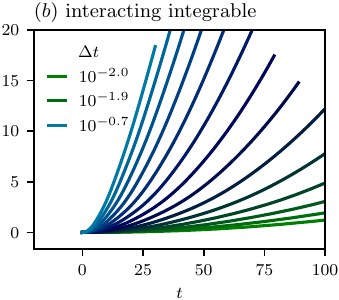}
\includegraphics{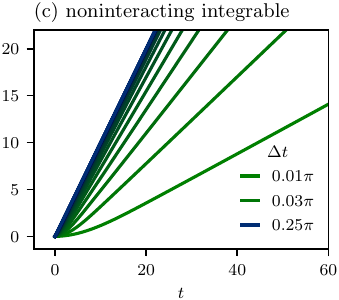}
    \caption{Krylov complexity $K(t)$ [Eq.~\eqref{eq:our_def_krylov_compl}] for the chaotic, interacting integrable, and noninteracting integrable at different values of the Trotter step $\Delta t$. In all cases we observe linear growth of the Krylov complexity for sufficiently large $t$. In chaotic and interacting integrable systems this linear growth has slope 1, whereas for noninteracting integrable systems this slope depends on the Trotter step and transitions to $1$ at $\Delta t =\pi/8$.}
    \label{fig:krylov_compl_growth}
\end{figure*}

\pwc{In this section we illustrate the growth of the Krylov complexity $K(t)$ for the three examples of circuits discussed in the previous section for completeness. Figure~\ref{fig:krylov_compl_growth} shows the growth of the Krylov complexity for different choices of Trotter step in (a) chaotic systems drawn from the GUE, (b) the interacting integrable XXZ model, and (c) the noninteracting integrable XX model. As we could already deduce from the analysis of the tridiagonal structure of the unitary superoperator $\mathcal{U}$, the Krylov complexity approaches a linear growth for large $t$ for the chaotic and interacting integrable systems. While there is no qualitative difference between the two, we observe that the Krylov complexity initially grows parametrically significantly slower in the integrable model, as also observed for Hamiltonian dynamics~\cite{parker_universal_2019,rabinovici_krylov_2022,rabinovici_krylovlocalization_2022}. Note that for small Trotter step many (discrete) time steps are necessary in order to approach this maximally ergodic regime, such that the crossover to the ergodic regime falls outside the considered range of time steps. This effect is especially pronounced in the integrable model due to the slower initial growth of Krylov complexity.}

\pwc{For the noninteracting integrable system we observe that, after some initial transient dynamics, the Krylov complexity also exhibits a linear growth. The slope of this growth now increases with increasing Trotter step and saturates for $\Delta t > \pi/8$. This behavior agrees with our analysis of the unitary superoperator and the discussed Trotter transition.}

\section{Conclusions and Discussion}
\label{sec:conclusions}
In this work we extended the notion of Krylov complexity to unitary circuit dynamics and applied this framework to study the effect of Trotterization, contrasting the Hamiltonian dynamics with the circuit dynamics.
During the unitary dynamics, an initial operator $O_0$ will spread out in a corresponding Krylov basis $\{\mathcal{O}_0, \mathcal{O}_1, \mathcal{O}_2,\dots\}$ obtained through a Gram-Schmidt orthonormalization procedure of the set of operators $\{O_0, U^{\dagger}O_0U, U^{\dagger2}O_0 U^2,\dots \}$. 
The autocorrelation function for $O_0$ only depends on its weight on $\mathcal{O}_0$ at any given time, giving the Krylov operators $\mathcal{O}_{n>0}$ the interpretation of `bath operators' that do not contribute to the autocorrelation dynamics.
In this basis the unitary evolution superoperator exhibits an upper Hessenberg matrix form that can be characterized with a limited number of parameters. 
We argued that for generic (chaotic) dynamics this superoperator attains a universal form after sufficiently many steps in the Krylov basis.

Physically, for generic dynamics the orthonormalization procedure underlying the Krylov approach converges to a set of operators that are here denoted `maximally ergodic Krylov operators'. 
These maximally ergodic operators present operators for which the unitary circuit dynamics is particularly simple: They are characterized by their instantaneously vanishing autocorrelation function, flat spectral function, and a maximal growth of Krylov complexity. For these operators the correlation function is hence indistinguishable from the random matrix prediction. 
In this sense the Gram-Schmidt orthonormalization procedure generically converges to maximally random bath operators.

In the specific case of unitary circuits obtained as a Suzuki-Trotter decomposition of local Hamiltonian dynamics, we observed that the approach to the maximally ergodic regime scales with the Trotter step: for smaller Trotter step the maximally ergodic operators, while still local, will have a larger support. For these operators the effect of Trotterization is most severe since the dynamics of their autocorrelation function is indistinguishable from purely random matrix dynamics. 

For integrable systems two different behaviors were observed. While interacting integrable systems also converged to the maximally ergodic regime, they did so on a (parametrically) longer scale and with the Lanczos coefficients exhibiting a noisier structure, as also observed in Hamiltonian dynamics. 
In noninteracting integrable systems the Krylov dynamics exhibited a nonanalytic transition as the Trotter step was varied.
For sufficiently short Trotter step the maximally ergodic regime is never reached, whereas after a critical Trotter step the Krylov dynamics again converges to the maximally ergodic regime. 

This transition can be interpreted as a transition in the conservation laws underlying integrability. For short Trotter steps these conservation laws remain sufficiently restrictive that they prevent maximally ergodic dynamics, as expected for integrable Hamiltonian dynamics, whereas at the critical Trotter step the effect of Trotterization becomes apparent and maximally ergodic Krylov dynamics is allowed. 
These nonanalytic Trotter transitions can be understood through an analytic calculation of the spectral function of the initial operator. For sufficiently short Trotter step this spectral function has a gap at the $\pi$-mode of the Floquet Brillouin zone, whereas this gap closes at the critical Trotter step. The orthonormalization procedure for operators translates to an orthonormalization procedure on the spectral function, which generally converges to a flat spectral function. However, the presence of a gap precludes the convergence to a flat spectral function and hence the convergence to maximally ergodic operators.

This work opens up various directions: Krylov complexity has been widely used in the recent study of Hamiltonian dynamics, and the presented framework can be directly applied to study unitary dynamics in the absence of any underlying static Hamiltonian, e.g., Floquet models. The correspondence between the spectral function and the Krylov dynamics strengthens the deep connection of spectral functions with quantum chaos. In Hamiltonian dynamics the spectral functions can exhibit different behavior for $\omega \to 0$ for chaotic and integrability-breaking vs. integrability-preserving dynamics \cite{pandey_adiabatic_2020,brenes_eigenstate_2020,leblond_universality_2021}, inviting similar studies using the Krylov approach. Here, the difference should be apparent in different limiting behaviors of the Krylov complexity. The connection between quantum chaos and delocalization in Krylov space should similarly still hold, and can be studied for more general unitary dynamics \cite{dymarsky_quantum_2020,rabinovici_krylov_2022,rabinovici_krylovlocalization_2022}. 
Analytical progress has recently been made for ETH by directly focusing on operator dynamics \cite{buca_unified_2023}, and a natural direction for future work would be to relate these results to Krylov dynamics. \pwc{While we focused on one-dimensional systems in this work, the developed framework also directly applies to higher-dimensional systems.}

Furthermore, it would be interesting to relate the notion of a bath in Krylov dynamics to similar notions of a bath in unitary circuit dynamics, such as in the influence matrix approach \cite{lerose_influence_2021,giudice_temporal_2022}. In a similar vein, Refs.~\cite{rakovszky_dissipation-assisted_2022,von_keyserlingk_operator_2022} proposed a classical simulation of operator dynamics by applying an artificial dissipation acting on long operators, allowing transport coefficients to be efficiently obtained. Backflow corrections from these long operators were argued to be exponentially small, similar to the corrections from the maximally ergodic operators in this work -- which also encode long operators --, presenting a further justification for this approximation. Exploring this connection would be useful for better understanding both the applicability of this classical approach and the emergence of hydrodynamics from operator dynamics.

The proposed approach may find applications in current Trotter-based methods to study quantum dynamics, both theoretical and analytical, since it highlights which operators are most sensitive to the effect of the Trotterization. The observed transition in noninteracting systems as the Trotter step is varied resembles the heating transition in periodically driven noninteracting systems~ \cite{prosen_nonequilibrium_2011,sen_entanglement_2016,tapias_arze_out--equilibrium_2020,aditya_dynamical_2022}, but crucially depends on the operator being studied, as apparent in the dependence of the critical Trotter step on the operator size. In this way unitary Krylov dynamics present a new probe to study Floquet transitions and unitary circuit dynamics.

\acknowledgments
The authors acknowledge useful discussions with Michael Flynn, Anatoli Polkovnikov, Ehud Altman, Michael Rampp, and Thomas Scaffidi. This work was in part funded by the Deutsche Forschungsgemeinschaft (DFG) via the cluster of excellence ct.qmat (EXC 2147, project-id 390858490).

\appendix 

\begin{comment}
\section{Measurement of Correlation function on quantum devices}
\label{sec:app_correlation_function_measurement}

{\color{blue}
The autocorrelation function $(O_0|O_t)$, Eq.~\eqref{eq:autocorrf} can be probed either directly on a quantum device by sampling from the $\mathrm{Tr}$ in the eigenbasis of the local operator $O_0$ yielding
\begin{align}
    \mathrm{Tr}\left[O_0^\dagger U^{\dagger t} O_0 U^t\right]/\mathcal{D} = \sum_n \lambda_n \langle n | U^{\dagger t} O_0 U^t |n\rangle/\mathcal{D}.
\end{align}
For instance for $O_0=\sigma^x$ we obtain $\lambda_n = \pm 1$, $|n\rangle = H|c\rangle$, where $|c\rangle$ is a state in the computational basis and $H$ the Hadamard gate. This way the autocorrelation function is measurable as a sum of an expectation value. As the initial state can be varied with any shot, the accuracy scales as $\sim 1/\sqrt{N}$, where $N$ is the number of shots. 
If one likes to avoid using the diagonal basis of $O_0$, one could also use an ancilla qubit like in \cite{Mi_2021}.
}
\end{comment}

\section{Unitary superoperator structure} \label{sec:app_superoperator_structure}

In this appendix we derive the structure of unitary superoperator $\mathcal{U}$ as stated in the main text [Eq.~\eqref{eq:U_nm}],
\begin{align}
&\mathcal{U}_{mn} = (\mathcal{O}_m|\,\mathcal{U}|\mathcal{O}_n) =\begin{cases}
0 \quad &\textrm{if} \quad m>n+1\,,\\
b_m \quad &\textrm{if} \quad m = n+1\,,\\
a_m\, c_n/c_m \quad &\textrm{if} \quad m<n+1\,,
\end{cases}\nonumber
\end{align}
with
\begin{align}
&a_n = (\mathcal{O}_n|\,\mathcal{U}|\mathcal{O}_{n}), \\
&b_n = (\mathcal{O}_{n}|\,\mathcal{U}|\mathcal{O}_{n-1}), \\
&c_n =  (\mathcal{O}_0|\,\mathcal{U}|\mathcal{O}_{n}).
\end{align}

We start by showing that Gram-Schmidt orthonormalization yields $(O_t|\mathcal{O}_n)=0$ for each $t<n$. In Eq.~\eqref{eq:decompose_ev_op_krylov} we define the lower triangular matrix $\alpha$ with elements $\alpha_{nt}$. As all $|\mathcal{O}_n)$ have to be linear independent, we need $\alpha_{nn}\neq 0$. Hence, the inverse matrix $\beta=\alpha^{-1}$ exists and is also a lower triangular matrix. We can write
\begin{align}
    |O_t) = \sum_{n=0}^t \beta_{t,n} |\mathcal{O}_n)\,,
\end{align}
as stated in Eq.~\eqref{eq:invert_transfo_krylov} in the main text. Using the orthonormality of $|\mathcal{O}_n)$, the statement $(O_t|\mathcal{O}_n)=0$ for $t<n$ follows immediately, since $|O_n)$ can be written as a linear combination of $|\mathcal{O}_{m \leq n})$. Note that $\beta$ is recursively the inverse of $\alpha$: The $l\times l$ reduced matrix $\beta$ is the inverse of the $l\times l$ reduced matrix $\alpha$. 

We hence find that $\mathcal{U}_{mn}$ is upper triangular, since for $m>n+1$,
\begin{align}
    \mathcal{U}_{mn}  =(\mathcal{O}_m|\mathcal{U}|\mathcal{O}_n) &= \sum_{ t=0}^{n} \alpha_{n,t} (\mathcal{O}_m|O_{t+1}) = 0\,,
\end{align}
using that $(\mathcal{O}_m|O_{t+1})=0$ since $t+1 \leq n+1 < m$. 

For $m \leq n$ we can write
\begin{align}
    \mathcal{U}_{mn} &= (\mathcal{O}_m|\,\mathcal{U}|\mathcal{O}_n) = \sum_{t=0}^{m} \alpha_{m,t}^* (O_t|\,\mathcal{U}|\mathcal{O}_n) \nonumber\\
    &= \alpha^*_{m,0} (O_0|\,\mathcal{U}|\mathcal{O}_n) + \sum_{ t=1}^{m} \alpha^*_{m,t} (O_{t}|\,\mathcal{U}|\mathcal{O}_n)  \nonumber\\
    & =\alpha^*_{m,0}\, \mathcal{U}_{0n} + \sum_{t=1}^{m} \alpha^*_{m,t} (O_{t-1}|\mathcal{O}_n)\nonumber\\
    &= \alpha^*_{m,0} \mathcal{U}_{0n} 
\end{align}
using the orthonormality $(O_{t-1}|\mathcal{O}_n)=0$ for $1 \leq t \leq n$ and that $|\mathcal{O}_0) = |O_0)$. As such, all matrix elements for $m \leq n$ factorize, where we can already identify $c_{n} = \mathcal{U}_{0n}$. For the remaining factor we can write $a_m = \mathcal{U}_{mm} =  \alpha^*_{m,0} \mathcal{U}_{0m} =   \alpha^*_{m,0} c_m$, resulting in the presented expression $\mathcal{U}_{mn}= a_m c_n/c_m$.

The remaining nontrivial matrix elements are those with $m = n+1$, which can not be simplified and which we denote as $b_n = \mathcal{U}_{n+1,n}$.

Due to the unitarity these sequences fulfill
\begin{align}
    1= \sum_{m=0}^{n+1} |\mathcal{U}_{m,n}|^2 = |b_n|^2 + |c_n|^2 \sum_{m=0}^{n} |a_m/c_m|^2
\end{align}
for all $n$, which we can rewrite as
\begin{align}
    1-|b_n|^2 = |c_n|^2 \sum_{m=0}^{n} |a_m/c_m|^2\,,
\end{align}
indicating how the lower diagonal elements are bounded by the remaining matrix elements.

\section{Orthonormal polynomials on the unit circle}
\label{app:orthonormalpoly}

In this appendix we review some properties of orthonormal polynomials on the unit circle that will serve to highlight the connection between the Krylov operators and the Fourier modes of the spectral function. Following Eq.~\eqref{eq:def_orthpoly}, there is a direct correspondence between the Krylov operators and polynomials on the unit circle $p_n(\omega)$, defined as
\begin{equation}
p_n(\omega) = \sum_{k=0}^n \alpha_{n,k}\, e^{ik\omega}\,,
\end{equation}
with the corresponding Krylov operator constructed as $|\mathcal{O}_n) = \sum_{k=0}^n \alpha_{n,k} |O_k)$. The orthonormality of the Krylov operators translates to the orthonormality of these functions provided we use the spectral function as weight function,
\begin{equation}
\int_0^{2\pi}\!\mathrm{d}\omega\, |f_O(\omega)|^2 p_n(\omega)p_m^*(\omega) = \delta_{mn}\,.
\end{equation}
While in the main text the Krylov operators were constructed through a Gram-Schmidt procedure, an explicit expression for these polynomials can also be obtained following standard results from, e.g., Ref.~\cite{szego_orthogonal_1975}. A similar approach for Krylov operators constructed from a Liouvillian was presented in Ref.~\cite{muck_krylov_2022}. Again writing the Fourier modes of the spectral function as
\begin{equation}
S_n =\int_0^{2\pi}\!\mathrm{d}\omega\, |f_O(\omega)|^2 e^{in \omega}\,,
\end{equation}
we consider Toeplitz matrices of the form
\begin{align}
    T_n = \begin{pmatrix}
        S_0 & S_{1} & S_{2} & \dots &\quad & S_{n} \\
        S_{-1} & S_{0} & S_{1} & \dots &\quad & S_{n-1} \\
        S_{-2} & S_{-1} & S_{0} & \dots &\quad & S_{n-2} \\
        \vdots & \vdots & \vdots & \ddots &\quad & \vdots \\
        S_{-n+1} & S_{-n+2} & S_{-n+3} & \dots &\quad & S_{1} \\
        S_{-n} & S_{-n+1} & S_{-n+2} & \dots &\quad & S_{0} \\
        \end{pmatrix},
\end{align}
where $(T_n)_{i,j} = S_{j-i}$. Since the spectral function has even parity we could also write $S_{n} = S_{-n}$, but the following results are more transparent by keeping the distinction between $S_n$ and $S_{-n}$.
The orthonormal polynomials can then be directly expressed in determinant form as
\begin{align}
    p_n(\omega) = \frac{1}{\sqrt{|T_n T_{n-1}|}} \begin{vmatrix}
    S_0 & S_{1} & S_{2} & \dots & \quad &S_{n} \\
        S_{-1} & S_{0} & S_{1} & \dots &\quad & S_{n-1} \\
        S_{-2} & S_{-1} & S_{0} & \dots &\quad & S_{n-2} \\
        \vdots & \vdots & \vdots & \ddots &\quad & \vdots \\
        S_{-n+1} & S_{-n+2} & S_{-n+3} & \dots &\quad & S_{1} \\
        1 & e^{i\omega} & e^{2i\omega} & \dots &\quad & e^{in\omega} \\
    \end{vmatrix}\,.
\end{align}
This expression is to be contrasted with the Krylov operators for Liouvillian dynamics, where the corresponding matrices are Hankel matrices composed out of the different moments of the spectral functions \cite{muck_krylov_2022}.

As such, all orthonormal polynomials can be expressed through Toeplitz matrices, and the expansion coefficients $\alpha_{n,k}$ can be analytically obtained as the corresponding minor from these matrices. The inverse transformation, writing 
\begin{align}
    e^{ik\omega} = \sum_{n = 0}^k\beta_{k,n} p_n(\omega)\,,
\end{align}
can also be obtained from a factorization of the Toeplitz matrices. Writing
\begin{align}
    &(T_n)_{k,\ell} = S_{\ell-k} =\int_0^{2\pi}\!\mathrm{d}\omega\, |f_O(\omega)|^2 e^{i(\ell-k) \omega} \nonumber\\
    &=\int_0^{2\pi}\!\mathrm{d}\omega\, |f_O(\omega)|^2\left(\sum_{m = 0}^k \beta^*_{k,m} p_m^*(\omega)\right) \left(\sum_{n = 0}^\ell \beta_{\ell,n} p_n(\omega)\right) \nonumber\\
    &= \sum_{m=0}^{\textrm{min}(k,\ell)} \beta^*_{k,m}\beta_{\ell,m}\,,
\end{align}
we can factorize the Toeplitz matrices as $T_n = B_n^{\dagger}B_n$, with $B_n$ an upper-triangular matrix defined as $(B_n)_{k,\ell} = \beta_{\ell,k}$ if $k \leq \ell$ and 0 otherwise.

Finally, we observe that the upper Hessenberg structure of the unitary superoperator also appears in the literature on orthonormal polynomials on the unit circle \cite{szego_orthogonal_1975}.

\begin{comment}
\section{Dual-Unitary Operators}\label{app:def_dual_unitary}
{\color{blue} In this section we state briefly the construction of a dual-unitary circuit for $q=2$, following \cite{bertini_exact_2019}. The circuit unitary $U$ is created by applying dual-unitary two qubit gates $U_{2j-1,2j}$ in a brickwork pattern. Each gate is drawn from the ensemble of gates
\begin{align}
V_{j,j'} &= e^{i \pi (\sigma^x_{j}\sigma^x_{j'}+\sigma^y_{j}\sigma^y_{j'})/4+iJ\sigma^x_{j}\sigma^x_{j'}} \\
U_{j,j'} &= u_{j} v_{j'} V_{j,j'} u'_{j} v'_{j'}
\end{align}
with $J$ uniformly distributed in $[0,2\pi)$ and $u,v,u',v'$ being Haar random single qubit gates. An explicit form for dual-unitary gates for larger $q$ is not required in this work.
}
\end{comment}

\section{Clifford Circuits}\label{app:Clifford}
In this appendix we consider operator dynamics generated by Clifford circuits where the initial operator is a Pauli operator. Clifford circuits have the property that they transform (direct) products of Pauli matrices to direct products of Pauli matrices. Since the product of Pauli matrices presents an orthonormal operator basis, for $|O_0)$ a Pauli matrix and Clifford dynamics it hence holds that $(O_s|O_t) = 0$ unless $O_s = O_t$.

\begin{figure}[h]
    \centering\includegraphics{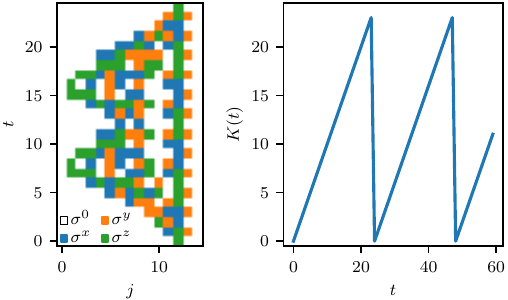}
    \caption{Clifford brickwork evolution for the initial operator $\sigma^z$ (left) and the corresponding Krylov complexity $K(t)$ (right). The circuit is generated by drawing $U_{j,j+1}$ at random from the set of all two-site Clifford unitaries.}
    \label{fig:krylov_compl_clifford}
\end{figure}

We illustrate the Krylov dynamics for a Clifford circuit in Fig.~\ref{fig:krylov_compl_clifford}. Each operator $U_{j,j+1}$ is randomly drawn from the set of all two-site Clifford unitaries~\cite{richter_transport_2023,corcoles_process_2013}. Every new generated operator $O_t$ is linear independent to all previous operators until the sequence repeats. The repetition is a consequence of the Floquet dynamics, using the same unitaries in each step: By drawing a random set of two-site Clifford gates, for some site $j$ $U_{j,j+1}$ commutes with all $O_t$, effectively reducing the space of accessible operators through localisation~\cite{chandran_semiclassical_2015}. Hence, for these cases, the Krylov complexity has a sawtooth shape, as shown in Fig.~\ref{fig:krylov_compl_clifford} for a typical example picked at random. The Krylov complexity initially grows with the maximal possible rate, until a revival occurs when the initial operator is recovered. The Krylov complexity then starts periodically repeating since the dynamics cycles through the different direct products of Pauli matrices generated by the Clifford circuit, which are exactly the orthonormal Krylov operators. This example also highlights how the dimension of the Krylov basis does not necessarily attain the maximal possible value, since the dimension is here set by the periodicity of the operator dynamics.

\section{Scaling Collapse}\label{sec:scaling_collapse}
\pwc{
In this appendix, we report on the details of the function $(\Delta t)^g$ used to obtain a scaling collapse in the main text. To be as general as possible, we allow $g$ to be $\Delta t$ dependent.
Then we obtain the function $(\Delta t)^g$ by interpolating the functions $a_n,b_n$ for different $\Delta t$ yielding $a(\Delta t,n)$ and $b(\Delta t,n)$ with $n \in \mathbb{R}$. In the next step, we iteratively compare $a(\Delta t,n)$ [or $b(\Delta t,n)$] at consecutive $\Delta t$. We start at the smallest $\Delta t = 10^{-2}$ and compare to the values at the next smallest $\Delta t = 10^{-1.9}$. We then obtain $g$ for this value of $\Delta t$ by minimizing
\begin{align}
   \int\!\mathrm{d}n\, |a(\Delta t,n)-a(\Delta t',n(\Delta t')^{g})|^2,
\end{align}
where we integrate over $n$ by interpolating the $a(\Delta t,n)$, or in the same manner for $b(\Delta t,n)$. This yields $g$ as a function of $\Delta t$. By extrapolation, we obtain $g$ for $\Delta t = 10^{-2}$ [GUE: $g(\Delta t=10^{-2})=1.43$, XXZ: $g(\Delta t=10^{-2})=2.1$]. This yields the results for $(\Delta t)^g$ depicted in Fig.~\ref{fig:results_gue_scaling_function}.
\begin{figure}[h]
    \centering
\includegraphics{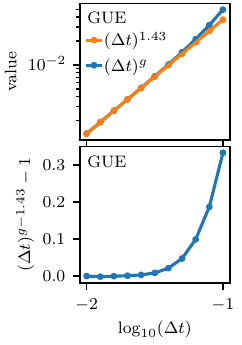}
\includegraphics{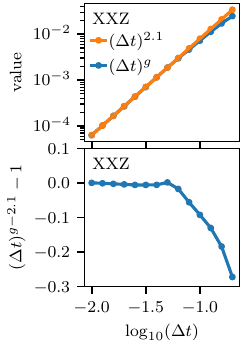}
    \caption{Rescaling factor $g$ for the rescaling function $(\Delta t)^g$ for the GUE system (left) and XXZ system (right), see Fig.~\ref{fig:tridiagonal_chaotic} and Fig.~\ref{fig:tridiagonal_xxz}, and comparison to $(\Delta t)^{1.43}$ and $(\Delta t)^{2.1}$, respectively.}
    \label{fig:results_gue_scaling_function}
\end{figure}
For small $\Delta t \rightarrow 0$, the scaling function seems to converge justifying the extrapolation to $\Delta t = 10^{-2}$, while deviations occur for large $\Delta t$. This may be expected, considering the operator spreading to converge for small $\Delta t$ (see Fig.~\ref{fig:operator_extent}), while it deviates for larger $\Delta t$.
}

\section{Convergence to the maximally ergodic Krylov regime}\label{app:convergence_large_n}
\begin{figure}
    \centering\includegraphics{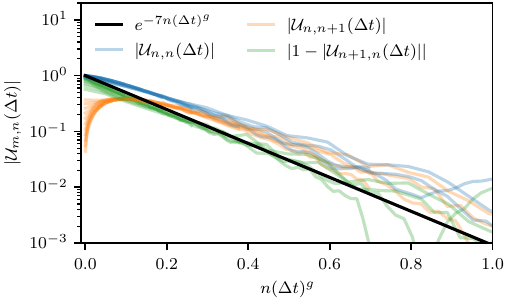}
    \caption{Convergence of the tridiagonal part of $\mathcal{U}$ to $\mathcal{U}_{n+1,n}=1,\mathcal{U}_{n,m}=0$ otherwise, with $U_{j,j+1}=\exp(-iH \Delta t)$ for a typical example of a Hamiltonian drawn from the GUE, the same as used in Fig.~\ref{fig:tridiagonal_chaotic}. The results were obtained using a system of size $70$ and TEBD with an MPS bond dimension of 1024 while varying $\Delta t$ from $10^{-2}$ to $10^{-1.2}$.}
    \label{fig:convergenve_towards_one_lower_diag_case_gue}
\end{figure}

In this appendix we consider the opposite limit, of large $\Delta t$ and $n$, highlighting how $\mathcal{U}$ approaches $\mathcal{U}_{n+1,n}=1$ exponentially. This convergence is shown in Fig.~\ref{fig:convergenve_towards_one_lower_diag_case_gue} for the chaotic system with the Hamiltonian drawn from the GUE ensemble. We note that for dual-unitary circuits and the XXZ Hamiltonian a similar exponential decay can be observed.

\section{Operator growth}\label{app:fit_error_function}
\begin{figure}[h]
    \centering
    \includegraphics{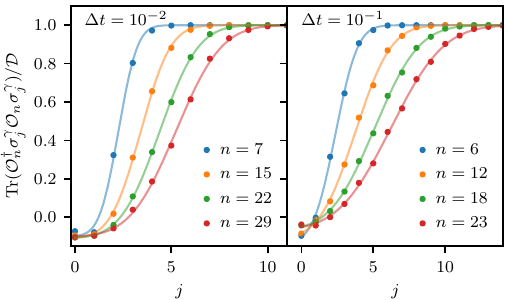}
    \caption{Fits (lines) to the OTOC data, (dots) Eq.~\eqref{eq:otoc_app}, using Eq.~\ref{eq:error_fct_app}. The OTOC data were generated using the Hamiltonian $H_{i,i+1}$ and the circuit structure also used for Fig.~\ref{fig:tridiagonal_chaotic} as well as the same numerical simulation settings.}
    \label{fig:fitting_error_fct}
\end{figure}

In this appendix we consider the operator growth of the Krylov operators $\mathcal{O}_n$. The operator growth is quantified through the OTOC
\begin{align}\label{eq:otoc_app}
\mathrm{OTOC}(n,j) = \left\langle \mathrm{Tr}\left[ \mathcal{O}_n^\dagger\sigma^\gamma_j \mathcal{O}_n \sigma^\gamma_j \right]\middle/\mathcal{D}\right\rangle_\gamma ,
\end{align}
where we averaged here over $\gamma \in \{x,y,z\}$, which we fit to an error function of the form
\begin{align}\label{eq:error_fct_app}
    f_{a,\ell,D}(j)=a+(1-a)\{\mathrm{erf}[ (j-\ell)/D ]+1\}/2\,.
\end{align}
The fitting results for exemplary $\Delta t,n$ are shown in Fig.~\ref{fig:fitting_error_fct}. We allowed the error function to deviate from $0$ by a value $a$ at $j=0$ to accommodate the observation that the OTOC is slightly negative at $j=0$. This signals that $\sigma^\gamma_0$ anticommutes rather than commute with $\mathcal{O}_n$. This may be a finite-time effect, and removing the fitting parameter $a$ does not change the results qualitatively: We still observe the $n^{0.65}$ growth as stated in the main text. 

\section{Convergence to the Hamiltonian regime}\label{sec:gue_convergence_hermtian}

In this appendix we report on the results for the convergence of the superoperator $\mathcal{U}$ towards the Liouvillian $\mathcal{L}$ in the limit $\Delta t\to 0$ for the cases where a chaotic Hamiltonian is drawn from the GUE and for the XXZ Hamiltonian. We consider the convergence of the diagonal elements $\mathcal{U}_{nn}=a_n$ towards $1$ and of the next-diagonal elements $\mathcal{U}_{n+1,n}=b_n$ towards the negative of $\mathcal{L}_{n+1,n}=\tilde{b}_n$, as expected in the limit $\Delta t\rightarrow 0$ from lowest order perturbation theory. 

\begin{figure}
    \centering
    \includegraphics{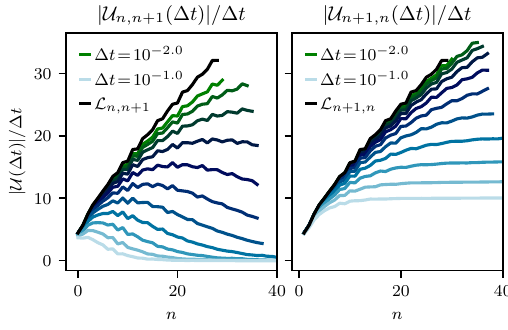}
    \includegraphics{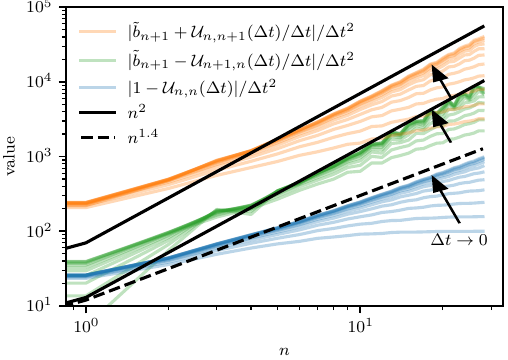}
    \caption{Convergence of the tridiagonal part of $\mathcal{U}$ to $\mathcal{L}$ with $U_{j,j+1}=\exp(-iH \Delta t)$ for a typical example of a Hamiltonian drawn from the GUE, the same used in Fig.~\ref{fig:tridiagonal_chaotic}. The results were obtained using a system of size $70$ and TEBD with an MPS bond dimension of $1024$ while varying $\Delta t$ from $10^{-2}$ to $10^{-1}$. Only numerically stable results are shown, determined by comparing to results obtained with a bond dimension of $512$.}
    \label{fig:convergenve_towards_hermitian_case_gue}
\end{figure}

For the GUE case, the convergence is shown in Fig.~\ref{fig:convergenve_towards_hermitian_case_gue}. Based on these results we expect $\mathcal{U}$ and $\mathcal{L}$ to return the same Krylov growth up to $n \sim 1/\Delta t$.

\begin{figure}
    \centering
    \includegraphics{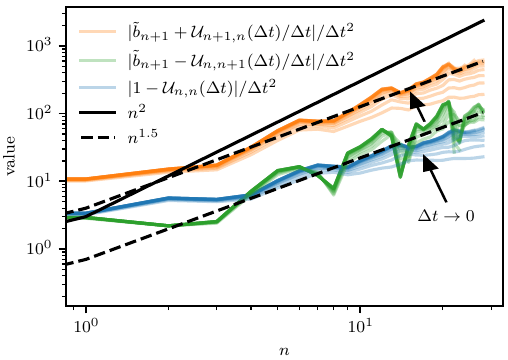}
    \caption{Convergence of the tridiagonal part of $\mathcal{U}$ to $\mathcal{L}$ for the interacting integrable XXZ Hamiltonian with $\Delta=3$. The results were obtained using a system of size $70$ and TEBD with an MPS bond dimension of $1024$ while varying $\Delta t$ from $10^{-2}$ to $10^{-0.7}$. Results for different $\Delta t$ are shown using opaque lines which converge for $\Delta t\rightarrow 0$ from below. Only numerically stable results are shown, determined by comparing to results obtained with a bond dimension of $512$.}
    \label{fig:convergenve_towards_hermitian_case_xxz}
\end{figure}

For the XXZ Hamiltonian, results are shown in Fig.~\ref{fig:convergenve_towards_hermitian_case_xxz}. It is interesting to note that for the XXZ Hamiltonian the scaling of the difference to $\mathcal{L}$ is different in comparison to the GUE example. Here, we observe the difference to grow as $\sim n^{1.4} \delta t^2$. Hence, for the XXZ circuit the superoperator $\mathcal{U}$ coincides with $\mathcal{L}$ up to $n \sim \Delta t^{-2/1.4}$, which is parametrically longer than for the GUE. 

\section{Detailed derivation for the integrable system} \label{app:detailed_der_integrable_system}

In the following we provide a detailed derivation for the exact solution of the integrable Trotterized XX model. 

We transform the XX Hamiltonian using the Jordan-Wigner transformation to obtain a description in terms of the fermionic operators $c_j,c_j^\dagger$ defined as
%%
%\pwc
{
\begin{align}
    \sigma^z_j &= 1-2 c_j^\dagger c_j, \\
    \sigma^x_j &= \prod_{l<j} (1-2 c_l^\dagger c_l) (c_j^\dagger +c_j),\\
    \sigma^y_j &= i\sigma^x_j \sigma^z_j =i\prod_{l<j} (1-2 c_l^\dagger c_l) (c_j^\dagger -c_j).
\end{align}
}
The Hamiltonian can hence be rewritten as
\begin{align}
    H_{j,j+1} &= \sigma^x_j \sigma^x_{j+1} + \sigma^y_j \sigma^y_{j+1} \nonumber\\
    &= (c_j^\dagger - c_j)(c_{j+1}^\dagger + c_{j+1}) - (c_j^\dagger + c_j)(c_{j+1}^\dagger - c_{j+1}) \nonumber\\
    &= 2 (c_j^\dagger c_{j+1} + c_{j+1}^\dagger c_j).
\end{align}
The unitary evolution operator is given by $U = e^{-iH_2 \Delta t}e^{-iH_1 \Delta t}$, where for the two layers of the time evolution we use
\begin{align}
    H_1 &= \sum_{j} H_{2j,2j+1}\\
    H_2 &= \sum_{j} H_{2j+1,2j+2}.
\end{align}
Both Hamiltonians preserve the momentum using a unit cell of size $2$, allowing us to introduce fermionic operators with fixed momentum as
\begin{align}
&c_{k+} = \frac{1}{\sqrt{N}}\sum_{j=0}^{N-1} e^{ikj} c_{2j}, \\
&c_{k-} = \frac{1}{\sqrt{N}}\sum_{j=0}^{N-1} e^{ikj} c_{2j+1},
\end{align}
where we consider a finite system with $N$ lattice sites, so that
\begin{align}
    H_1 &= 2\sum_{k} c_{k+}^\dagger c_{k-} + c_{k-}^\dagger c_{k+}, \\
    H_2 &= 2\sum_{k} e^{-ik} c_{k+}^\dagger c_{k-} + e^{ik} c_{k-}^\dagger c_{k+}\,.
\end{align}
Now we can use that
\begin{align}
    [H_1,c_{k+}] &= -2c_{k-}(c_{k+}^\dagger c_{k+} +c_{k+} c_{k+}^\dagger) = -2c_{k-}\,, \\
    [H_1,c_{k+}^\dagger] &= 2c_{k-}^\dagger (c_{k+} c_{k+}^\dagger+c_{k+}^\dagger c_{k+}) = 2c_{k-}^\dagger\,,
\end{align}
and similarly
\begin{align}
    [H_1,c_{k-}] &= -2c_{k+}, &&\quad
    [H_1,c_{k-}^\dagger] &&= 2c_{k+}^\dagger, \\
    [H_2,c_{k+}] &= -2c_{k-} e^{-ik}, &&\quad
    [H_2,c_{k+}^\dagger] &&= 2c_{k-}^\dagger e^{ik},\\
    [H_2,c_{k-}] &= -2c_{k+}e^{ik}, &&\quad
    [H_2,c_{k-}^\dagger] &&= 2c_{k+}^\dagger e^{-ik}.
\end{align}

The time evolution for $H_{1}$ now follows from
\begin{align}
     \frac{\mathrm{d}}{\mathrm{d}t} e^{i H_1 t} \begin{pmatrix} c_{k+} \\ c_{k-} \end{pmatrix} e^{-i H_1 t}   = - i \begin{pmatrix} 0 & 2 \\ 2 & 0 \end{pmatrix} e^{i H_1 t} \begin{pmatrix} c_{k+} \\ c_{k-} \end{pmatrix} e^{-i H_1 t} \,
\end{align}
so that we can write
\begin{align}
    e^{i H_1 \Delta t} \begin{pmatrix} c_{k+} \\ c_{k-} \end{pmatrix} e^{-i H_1 \Delta t}   = \begin{pmatrix}\cos(2 \Delta t) & -i\sin(2 \Delta t) \\ -i\sin(2 \Delta t) & \cos(2 \Delta t) \end{pmatrix} \begin{pmatrix} c_{k+} \\ c_{k-} \end{pmatrix}.
\end{align}
For $H_2$ we similarly find
\begin{align}
     \frac{\mathrm{d}}{\mathrm{d}t} e^{i H_2 t} \begin{pmatrix} c_{k+} \\ c_{k-} \end{pmatrix} e^{-i H_2 t}   = - i \begin{pmatrix} 0 & 2e^{-ik} \\ 2e^{ik} & 0 \end{pmatrix} e^{i H_1 t} \begin{pmatrix} c_{k+} \\ c_{k-} \end{pmatrix} e^{-i H_1 t}\,.
\end{align}
\begin{widetext}
Using the eigenvectors $(1,e^{ik})^T/\sqrt{2}$, $(-e^{-ik},1)^T/\sqrt{2}$ with eigenvalues $2,-2$, we get
\begin{align}
    e^{i H_2 \Delta t} \begin{pmatrix} c_{k+} \\ c_{k-} \end{pmatrix} e^{-i H_2 \Delta t}   &= \frac{1}{2}\left[ \begin{pmatrix}1 & e^{-ik} \\ e^{ik} & 1\end{pmatrix}e^{-2i\Delta t} \right.\left.+\begin{pmatrix}1 & -e^{-ik} \\ -e^{ik} & 1\end{pmatrix}e^{2i\Delta t} \right] \begin{pmatrix} c_{k+} \\ c_{k-} \end{pmatrix} =\begin{pmatrix}\cos(2 \Delta t )& -i\sin(2 \Delta t )e^{-ik} \\ -i\sin(2 \Delta t )e^{ik}& \cos(2 \Delta t )\end{pmatrix} \begin{pmatrix} c_{k+} \\ c_{k-} \end{pmatrix}.
\end{align}
We have to multiply the matrix representations of the two layers to obtain
\begin{align}\label{eq:app:U_nonint}
    \mathcal{U}_k &= \begin{pmatrix}\cos(2 \Delta t )& -i\sin(2 \Delta t )e^{-ik} \\ -i\sin(2 \Delta t )e^{ik}& \cos(2 \Delta t )\end{pmatrix}
    \begin{pmatrix}\cos(2 \Delta t) & -i\sin(2 \Delta t) \\ -i\sin(2 \Delta t) & \cos(2 \Delta t) \end{pmatrix} \\
    &= \begin{pmatrix}1-\sin(2 \Delta t)^2(1+e^{-ik}) & -i\cos(2\Delta t)\sin(2 \Delta t)(1+e^{-ik}) \\ -i\cos(2\Delta t)\sin(2 \Delta t)(1+e^{ik}) & 1-\sin(2 \Delta t)^2(1+e^{ik}) \end{pmatrix}.
\end{align}
This returns the expression from the main text. The matrix can be directly diagonalized to return eigenvalues $e^{\pm i \omega_k}$, with
\begin{align}
\cos(\omega_k) =1 - 2 \sin^2(2\Delta t)\cos^2(k/2)\,.
\end{align}
The expressions for the autocorrelation function \eqref{eq:corr_c0} then follow by making use of the fact that
\begin{align}
    (O_0|k+) = 1, \qquad (O_0|k-) = 0\,,
\end{align}
and expressing $|O_0)$ in the eigenbasis of Eq.~\eqref{eq:app:U_nonint}.
\end{widetext}

\end{document}